\documentclass[referee,a4paper,12pt,traditabstract]{jswsc} 

\usepackage{graphicx}
\usepackage{txfonts}
\usepackage{subfigure}
\usepackage{epstopdf}
\usepackage[displaymath,mathlines]{lineno}
\usepackage[authoryear,round]{natbib}
\usepackage[backref]{hyperref}
\usepackage{url}
\usepackage{adjustbox}
\usepackage{float}
\usepackage{makecell}

\hypersetup{colorlinks=true,citecolor=cyan,urlcolor=cyan,linkcolor=blue}

\begin{document}

% \begin{linenumbers}  

   \title{A catalogue of observed geo-effective CME/ICME characteristics}

   % \subtitle{Extension of an existing catalogue using propagation parameters derived from Drag-Based Model}
   
   \titlerunning{DBM inversion Procedure}

   % \authorrunning{Wuchterl and Ptolemy}

   \author{R. Mugatwala\inst{1,2}\and
           S. Chierichini\inst{2,1}\and
           G. Francisco\inst{1,3}\and
           G. Napoletano\inst{4,1}\and
           R. Foldes\inst{4,5}\and
           L. Giovannelli\inst{1}\and
           G. De~Gasperis\inst{6}\and
           E. Camporeale\inst{7,8}\and
           R. Erd\'elyi\inst{2,9,10}\and
           D.~Del~Moro\inst{1}
          \fnmsep
          }

   \institute{Department of Physics, University of Rome ``Tor Vergata'', Rome, Italy\\
              \email{\href{mailto:ronish@roma2.infn.it}{ronish@roma2.infn.it}}
              \thanks{Corresponding Author}
         \and
             SP2RC, School of Mathematics and Statistics, University of Sheffield, Sheffield, England
        \and
             Institute of Astrophysics and Space Science, University of Coimbra, Coimbra, Portugal  
        \and
             Department of Physical and Chemical Sciences, University of L'Aquila, L'Aquila, Italy
        \and
            Laboratoire de Mecanique des Fluides et d'Acostique, CNRS, Universite Claude Bernard Lyon
        \and Dipartimento di Fisica, Sapienza Universit\`a di Roma, P. le A. Moro 2, Roma, Italy  
        \and
             CIRES, University of Colorado, Boulder, USA
        \and    
             NOAA Space Weather Prediction Center, Boulder, USA 
        \and
             Department of Astronomy, E\"otv\"os Loránd University, Budapest, Hungary
        \and 
            Gyula Bay Zolt\'an Solar Observatory (GSO), Hungarian Solar Physics Foundation (HSPF), Gyula, Hungary
            }

%%   \date{Received September 15, 1996; accepted March 16, 1997}

  % \abstract{}{}{}{}{}        %% uncomment if structured abstract is desired
 %% 5 {} token are mandatory
 
  \abstract
 %% context heading (optional). leave {} empty if necessary  
   { One of the goals of Space Weather studies is to achieve a better understanding of impulsive phenomena, such as Coronal Mass Ejections (CMEs), in order to improve our ability to forecast them and mitigate the risk to our technologically driven society. 
   % To achieve this, it is crucial to assess the performance of forecasting models. 
   The essential part of achieving this goal is to assess the performance of forecasting models. 
   To this end, the quality and availability of suitable data are of paramount importance. 
   In this work, we have merged already publicly available data of CMEs from both in-situ and remote instrumentation in order to build a database of CME properties. 
   To evaluate the accuracy of such a database and confirm the relationship between in-situ and remote observations, we have employed the drag-based model (DBM) due to its simplicity and inexpensive cost of computational resources. 
   %DBM is an analytical model that assumes the aerodynamic drag caused by the surrounding solar wind to be the primary factor in the interplanetary propagation of CMEs. 
   In this study, we have also explored the parameter space for the drag parameter and solar wind speed using a Monte Carlo approach to evaluate how well the DBM determines the propagation of CMEs for the events in the dataset. 
   % The dataset of geoeffective CMEs constructed as a result of this work provides validation of the initial hypothesis about DBM, solar wind speed and also yields additional information about CMEs like arrival time, arrival speed, lift-off time, etc and also provides statistical measurements of the DBM model parameters. 
   The dataset of geoeffective CMEs constructed as a result of this work provides validation of the initial hypothesis about DBM, and solar wind speed and also yields further insight into CME features like arrival time, arrival speed, lift-off time, etc.
   Furthermore, the dataset also provides statistical metrics for the DBM model parameters.  
   Also, the probability distribution function for the free parameters of DBM has been derived through a Monte Carlo-like inversion procedure.
   Probability functions obtained from this work are comparable to distributions employed in previous works.
   Using a data-driven approach, this procedure allows us to present a homogeneous, reliable, and robust dataset for the investigation of CME propagation.
   On the other hand, possible CME events are identified where DBM approximation is not valid due to model limitations and higher uncertainties in the input parameters, those events require more thorough investigation.}        %% replace by pair of curly brackets, {}, if structured abstract is selected


   \maketitle
%%
%%________________________________________________________________

\section{Introduction}
%
%Coronal Mass Ejections (CMEs) are huge clouds of magnetic field and plasma released from the Sun. The interplanetary part of these CMEs called ICMEs and those ICMEs are responsible for major geomagnetic storms \citep{Koskinen2007}. A Southward pointing IMF ($B_Z$ field) of ICMEs may lead to well-known consequences on Earth's magnetosphere and as a result space and ground-based technologies are threatened \citep{Gosling1991, Tsurutani1988, Schwenn2006a, Temmer2021}. Thus, it's essential to develop and advance our forecast methods in order to get reliable predictions for Time of Arrival (ToA) and Speed of Arrival (SoA) of CMEs. In the last two decades, a lot of efforts have been made by the space scientists community in order to get accurate predictions but it's not an easy task \citep{Riley2018,Vourlidas2019}.    
%
ICMEs (Interplanetary Coronal Mass Ejections) are eruptions of plasma and magnetic fields from the Sun's corona that propagate in the  Heliosphere \citep{webb2012coronal}.
These plasma and magnetic field structures ejected from the Sun travel through the interplanetary space environment and reach the 1 AU range within 1-5 days \citep{chen2011coronal}. 
In in-situ data, ICMEs can be discerned from the average solar wind by their distinct signatures, such as the enhanced magnetic field, the higher particle speed, and the variations in plasma density \citep{liu2010reconstructing, papaioannou2016solar}.
They can also be observed remotely by using instruments such as coronagraphs (particularly SOHO/LASCO with coronagraphs C1/C2/C3 \citet{Domingo+1995, brueckner1995large}, STEREO/SECCHI with COR1/COR2 \citet{Kaiser+2008, howard2008sun} ), and Heliographic imagers (HI1/HI2) \citep{eyles2009heliospheric}.

ICMEs are among the main drivers of Space Weather, impacting the space environment and human technology \citep{Gosling1991, Tsurutani1988, Schwenn2006a, pulkkinen2007space, Temmer2021}. 
The plasma and magnetic fields ejected from the Sun can interact with Earth's magnetic field, leading to geospace disturbances \citep{Koskinen2007}, which affect a wide range of technological systems in space, such as satellites, telecommunications, and the GNSS systems \citep{shea1998space, schrijver2010heliophysics, aquino2013correlation, piersanti2017comprehensive}. 
The present strategies to mitigate the effects of ICMEs on space-based technologies and infrastructures require the knowledge of the ICME arrival time with low uncertainty to allow operators to take action to protect their equipment, by shutting them down or putting them in a safe mode \citep{barbieri2004october, sreeja2016impact, veettil2019ionosphere}. 

In the last decades, space agencies have designed and launched missions to observe the Sun and monitor the solar wind characteristics, and track CMEs and ICMEs as they travel through space, with the aim to study their interactions with the interplanetary environment.
%
%Still, the characteristics of ICMEs are difficult to predict \citep{manchester2017physical, Riley2018, Vourlidas2019}, and the scientific community is approaching the issue of forecasting their Time-of-Arrival (ToA) and Speed-at-Arrival (SaA) at Earth, and the magnitude and the sign of the southward component ($Bz$) of the associated magnetic field, this latter deemed the most important parameter for the strength of the resulting geomagnetic storm.
Despite these advancements in space weather forecasting, accurately predicting the characteristics of ICMEs such as their Time-of-Arrival (ToA) and Speed-at-Arrival (SaA) at Earth, as well as the magnitude and direction of the southward component of their magnetic field \citep[which is crucial for determining the intensity of geomagnetic storms][]{koskinen2006geoeffectivity}, remains a challenging task for the scientific community \citep{manchester2017physical, Riley2018, Vourlidas2019}.

Following the evolution of numerical methods and the increase of available computational power, a number of empirical methods, physics-based analytical models, and MHD numerical simulations for the ICME kinematics have been developed.
In the MHD approximation, the boundary conditions are derived from observed magnetograms and coronographic images and model the propagation of the ejecta by numerically solving the magneto-hydrodynamic equations \citep[ENLIL, HAFv.2 (Hakamada-Akasofu-Fry version 2)+3DMHD, EUHFORIA (EUropean Heliospheric FORecasting Information Asset)][]{odstrcil2003numerical, wu2007three, pomoell2018euhforia}.
%These simulations make it possible to account for the physical processes that we model, of course, their use requires expensive computing resources.
These simulations allow for the inclusion and consideration of the physical processes being modelled.
However, their use requires substantial computing resources due to their computationally intensive nature, making them expensive to run.
%While, as stated by \citet{manchester2017physical} “the future of space weather forecasting lies in numerical modelling”, it is also true that currently “the performance of empirical and analytical methods is comparable to, or even slightly better than that of numerical methods”.\\
%
The complete understanding of the physical processes involved in the Sun-Earth relation relies heavily on numerical modelling techniques.
However, with present observation capabilities, the forecasting performance of empirical and analytical methods are comparable to, or in some cases slightly better than, those achieved with numerical methods due to uncertainties in the input parameters \citep{manchester2017physical}. 
This implies that the existing empirical and analytical approaches are still effective and competitive in terms of their predictive capabilities and that the near future of space weather forecasting lies with the use of these computationally light approaches and Machine Learning (ML). \\ \\
In general, analytical methods are computationally lighter and their parameters can be easily updated with new incoming data.
Also, physics-based analytical models \cite[e.g.,][]{Vrvsnak2013DBM,rollett2016elevohi, paouris2017effective, Napoletano2018} can shed light on the ICME dynamics, and this knowledge would possibly help us in refining also numerical methods.
On the other hand, the relationships between ToA and SaA and various CME parameters measured at (or close to) their launch, have been used in empirical prediction methods \citep[e.g.,][]{manoharan2006evolution, gopalswamy2009coronal}, and most recently in a plethora of ML approaches.
ML techniques have become more and more used in space weather, as recently reviewed in \citet{camporeale2019challenge}. 
In the last years, there have been many attempts to leverage on ML algorithms to obtain the characteristics of an ICME at L1 from the associated CME observables \citep[ just to list a few]{bobra2016predicting, liu2018new, wang2019new}.
These ML algorithms use catalogues of CME/ICME characteristics for the training, in order to set their parameters, validate their results and check their performances.
Consequently, it becomes more and more important to build CME/ICME databases with a large number of events and small uncertainties \citep[ML methods typically need numerous, relevant and reliable examples in the datasets in order to give accurate results][]{vanderplas2012introduction,  ivezic2014statistics}. \\\\
%In this paper, we propose a method to update the catalogue of CME-ICME pairs presented in \citet{Napoletano2022}, by using a Monte Carlo approach to validate its entries.
%Then, we make use of this refined catalogue to infer the Probability Distribution Functions (PDFs) to use for the P-DBM method \citep{Napoletano2018, del2019forecasting}.
In this paper, we present a method to update the catalogue of CME-ICME pairs published in \citet{Napoletano2022}, by using a constrained Monte-Carlo strategy to validate its entries.
The constrained Monte Carlo strategy allowed us to explore the parameter space in a more effective way.
We then make use of this updated catalogue to revisit the Probability Distribution Functions (PDFs) to use for the P-DBM method \citep{Napoletano2018, del2019forecasting}.
Finally, we present a comparison of these PDFs for different solar wind conditions and against previous literature.\\
The paper is organized as follows. Section \ref{METHODS} describes the DBM model and the mathematical methodology to retrieve PDFs from the catalogue.
In Section \ref{RESULTS}, we analyse the results of the inversion and use them to relabel the CME/ICME catalogue entries and to obtain PDFs for different ICME types.
Section \ref{DISCUSSION} is dedicated to conclusions and discussions.\\
The CME-ICME dataset compiled and used in this work can be found at \url{https://zenodo.org/record/8063404} and a description of the different column headers is provided in the appendix \ref{Dataset Description}.  
   
\section{Methods}
\label{METHODS}
\subsection{Drag-Based Model (DBM)}
The Drag-Based Model is one of the simplest models that describes CME propagation through the heliosphere. Due to its simplicity and calculation speed, it is one of the most popular models used in CME forecast tools. In recent years, DBM has been used in many studies to describe CME propagation which are summarised in \cite{Dumbovic2021}.\\
DBM is based on the assumption that the responsible Lorentz force for CME launch is negligible in the upper part of the solar corona (after a certain heliocentric distance of 20 R$\odot$, but this assumption is not always valid as for many events Lorentz force is still comparable with drag force and upper limit of the heliocentric distance vary from event to event \citet{vrvsnak2001dynamics, vrvsnak2004kinematics, Sachdeva2015, Sachdeva2017} ) and beyond this heliocentric distance the dynamics of the ICME is dominantly governed by its interaction with ambient solar wind via MHD drag \citep{Cargill2004, Vrvsnak2013DBM}. 
Due to MHD drag force ICMEs that are faster (slower) than solar wind have a tendency to decelerate (accelerate) during propagation, which was also supported by observations \citep{Gopalswamy2000}.
CME radial acceleration according to the DBM approach is given as: 
\begin{equation}
\label{eqn:acceleration}
        a(r) = - \gamma  \bigg(v(r)_{CME}-w\bigg)  \mid v(r)_{CME}-w \mid
\end{equation}
where $a(r)$ and $v(r)_{CME}$ are the instantaneous acceleration and speed of ICME, respectively, $w$ is the instantaneous ambient solar wind speed, $\gamma$ is the drag parameter that is also called as drag efficiency.
It is important to note that all the quantities in equation \ref{eqn:acceleration} are space and time-dependent.
Also, beyond 20 R$\odot$, $\gamma$ and $w$ may be approximated to be constant throughout the heliosphere \citep{Cargill2004, Vrvsnak2013DBM}. Under such approximation, equation \ref{eqn:acceleration} can be solved analytically to obtain heliospheric distance and speed of ICME as a function of time \citep{Vrvsnak2013DBM}:
\begin{equation}  
 v(t) = \frac{v_0 - w}{1 \pm \gamma (v_0 - w)t}+w
\label{eqn:vel}
\end{equation}

\begin{equation}
r(t)= \pm \frac{1}{\gamma} \ln(1 \pm \gamma (v_{0}-w)t) + wt +r_{0}
\label{eqn:dis}
\end{equation}
where $\pm$ sign accounts for accelerated/decelerated CMEs i.e., plus for v$_{0} >$ w and minus for v$_{0}<$ w.
Eqns. \ref{eqn:vel} and \ref{eqn:dis} give us the speed and distance as a function of CME propagation time from an initial distance (at t=0) r$_{0}$ and take-off speed v$_{0}$.
From those, one can determine the transit time $t_{1AU}$ and impact speed $v_{1AU}$ at 1AU.   %transit time T needed for ICME to travel from an initial distance (t=0) r$_{0}$ to any defined distance $r_1$ and also gives the impact speed at a distance $r_1$ using initial take-off speed v$_{0}$ at t=0 \citep{vrvsnak2013propagation}.

\subsection{DBM Inversion Procedure} \label{inversion}
DBM solution, as given in \cite{Vrvsnak2013DBM}, can be used to obtain the analytical values of free DBM parameters.
If the ICME follows the DBM model, and if its boundary conditions, i.e, initial position $r_0$, initial speed $v_0$, ToA $t_{1AU}$ and impact speed $v_{1AU}$ are known, then the free parameters of the model, namely drag parameter $\gamma$ and solar wind speed $w$, can be obtained via a mathematical inversion of the set of equations presented above (eqs. \ref{eqn:vel} and \ref{eqn:dis}).  %two equations \ref{eqn:dis} and \ref{eqn:vel} \cite{Napoletano2018}.

\begin{equation}
    \frac{(v_0 - w)(v_{1AU} - w)t_{1AU}}{(v_0 - v_{1AU})} \ln \left[\frac{(v_0 - v_{1AU})}{(v_{1AU} - w)}+1\right] + w t_{1AU}+ r_0 - r_{1AU} = 0 
    \label{inv-1}
\end{equation}
 The above equation \ref{inv-1} is solved numerically to obtain $w$, then using equation \ref{inv-2} is used to directly compute $\gamma$:
 
 \begin{equation}
     \gamma =\frac{(v_0 - v_{1AU})}{(v_0 - w)(v_{1AU} - w)t_{1AU}}
     \label{inv-2}
 \end{equation}

 \subsection{Mathematical Framework}
In search for the unique distribution for the free DBM parameters, we applied the DBM inversion procedure to the existing dataset, published in the previous works of \cite{Napoletano2018} and \cite{Napoletano2022}.
A comprehensive description of a dataset is provided in appendix \ref{Dataset Description}, while the summary of a few particular quantities used in this study and associated results are tabulated in table \ref{dataset}. 
In the process of DBM inversion, we discovered that the majority of the CME events in the dataset lack analytical solutions for the equations \ref{inv-1} and \ref{inv-2}.
%need a refinement for next line.
This is unexpected, since "DBM is (reasonably) valid during CME propagation for all the CME events" is a null hypothesis for the dataset. 
The reason behind this discrepancy is that errors associated with the initial position ($r_0$), target position ($r_1$), transit time (t$_{1AU}$), impact speed ($v_1$) and initial speed ($v_0$) were not taken into account in the inversion procedure.
Another possible reason is that DBM is not properly describing the CME motion (e.g, $w =$ constant is not a realistic approximation; CME-CME interaction is also possible). 
\par
However, for this work, we adhere to our null hypothesis and consider the possibility of including uncertainties for the measured CME features.
To incorporate the errors associated with those quantities, we adapted a pairwise selection approach. It is worth noting that \cite{Napoletano2022} also adopted a probabilistic approach in the inversion procedure to obtain $w$ and $\gamma$. 
In order to do that, they assumed that [$r_0,r_1,v_0,v_1,t_{1AU}$] follows a normal distribution and draw random samples, here the majority of samples are concentrated in the part of the Gaussian curve peak.
However, our pairwise approach allowed us to explore other parts of parameter space where less probable values exist. 
We have assumed that two parameters, $r_0$ and $r_1$, do not suffer any errors because their values are fixed. We took $r_0$ = 20 R$\odot$ and, for $r_1$ we have used the actual Sun-Earth Distance at a time when CME is at $r_0$. 
% We have also assumed that the arrival speed has no intrinsic errors as there is no associated entry in the database.
The arrival speed of CME in the dataset is calculated as a mean of solar wind speed during a disturbance in plasma and therefore it has associated intrinsic error.
The error associated with arrival speed is relatively small compared to the initial speed and arrival time, therefore it is neglected in the study.
Therefore, we end up with only two quantities, $v_0$ and $t_{1AU}$ which have larger errors.
Next, we made a pairwise selection of ($t_{1AU},v_0$) for each DBM inversion iteration from the normal distribution followed by both quantities where $\mu$ is the observed value ("Transit\_Time" and "v\_r" is taken as $\mu$) and $\sigma$ is an error associated with the observed quantity ("Transit\_time\_err" and "v\_r\_err" taken as $\sigma$).
It is important to keep in mind that the tails of the normal distribution function are 3$\sigma$ width.
For a pairwise selection, we draw 200 samples for t$_{1AU}$ and the same for $v_0$. So in the end, we have a total of 40,000 possible pairs. 
After this pair selection, we performed the DBM inversion to obtain values of $w$ and $\gamma$, respectively. 
\par The DBM inversion procedure is a Monte Carlo process and after the inversion procedure, we have 40,000 possible solutions for $w$ and $\gamma$.
Many of these values can not be physically feasible, for example, negative values of $w$.
\cite{Vrvsnak2013DBM} provided a brief description of their work, and from there we deduced that the drag parameter $\gamma$ has a relation with the mass of the CME. 
In our primary analysis, we found that the inversion procedure also provides very high values for $\gamma$ which can not be explained by the typical mass range of CMEs.
Therefore, it is necessary to employ constraints on the values obtained through the inversion procedure. The constraints that we imposed on inversion values are given below.
\begin{enumerate}
    \item 0 $\leq$ $w$ $\leq$ $1000$ km/s\\
    Solar wind speed cannot be negative and the typical speed for fast solar wind in literature is $800 km/s$. It is worth noting that the condition of realistic solar wind speed in \cite{Paouris2021a} is 300-600 $km/s$ which is very narrow compared to us. 
    \item $0.1\times 10^{-7}$ $\leq$ $\gamma$ $\leq$ $3.0\times10^{-7}$ km$^{-1}$ \\
    It is important to note that the typical range for the $\gamma$ parameter, in \cite{Vrvsnak2013DBM}, is $0.2 - 2.0 \times 10^{-7}$, but we widen this range to accept a few more extreme solutions. Similarly, \cite{Paouris2021a} has a range of $0.01 -0.59 \times 10^{-7}$ for realistic drag parameter but their obtained values are in the range of $0.21 - 0.42 \times 10^{-7}$ (see table-4 of \citep{Paouris2021a}) which is comparable to our range. 
\end{enumerate}
 
\begin{figure}[h]
    \centering
    \includegraphics[width=\textwidth]{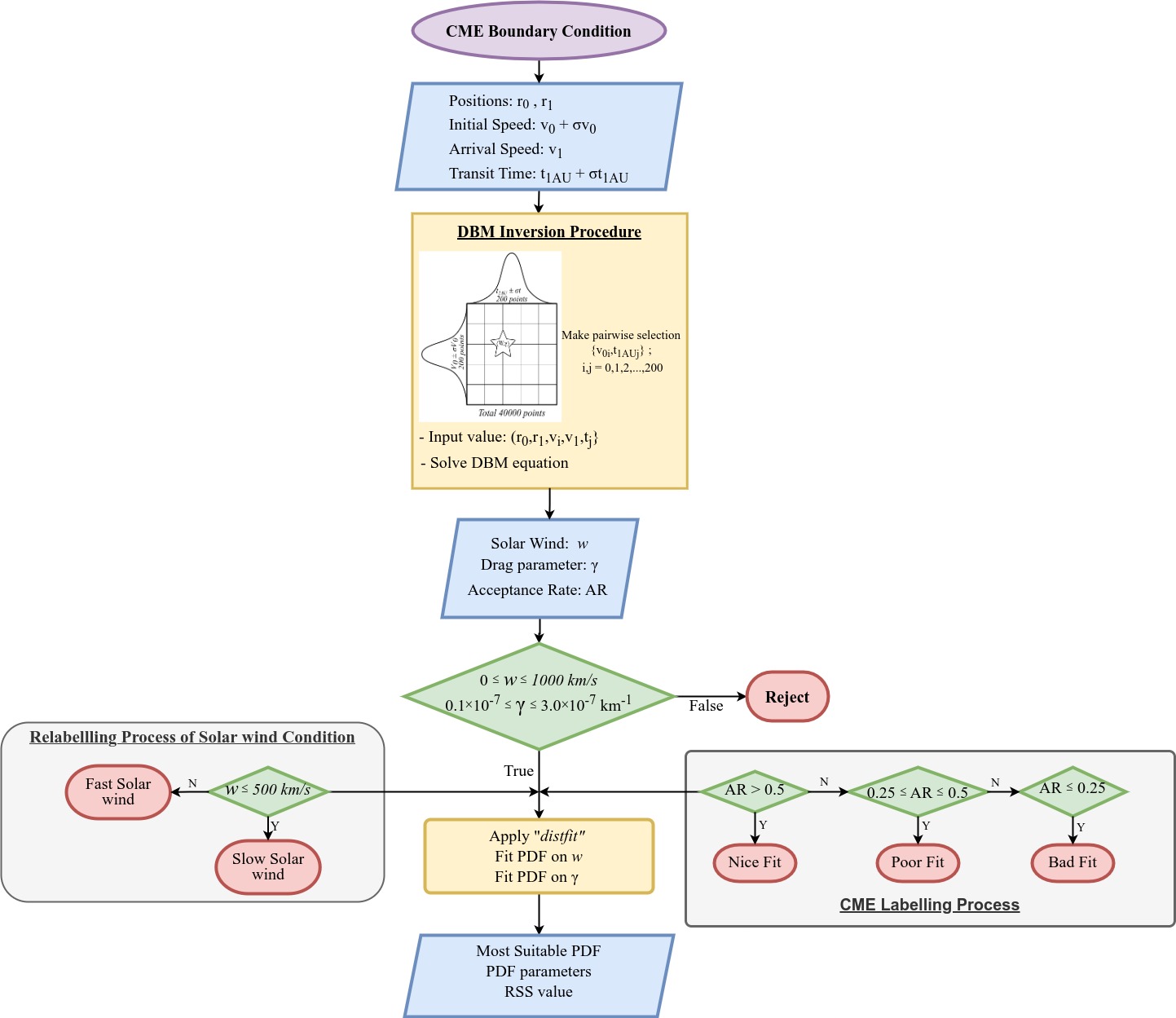}
    \caption{Schematic of DBM inversion procedure. The values of boundary conditions are fed into the equations \ref{inv-1}  and \ref{inv-2} using a pairwise approach to obtain $w$ and $\gamma$. The obtained values are checked for selection criteria. The accepted values are used to determine the solar wind condition, the most suitable PDF of model parameters and CME labelling scheme.  }
    \label{fig:flowchart}
\end{figure}

After this, we derive the four main quantities namely $W_{mean}$, $\gamma_{mean}$, $W_{opt}$ and $\gamma_{opt}$ from the accepted values of $w$ and $\gamma$; the $opt$ values correspond to the DBM input that produced the minimum deviation from the observed transit time.
In order to evaluate the "goodness" of the inversion procedure, we define the \textit{"Acceptance Rate"} as the ratio between the number of meaningful solutions to the total number of possible solutions, represented by the number of samples.
\begin{equation}
\label{eqn:AR}
    \textrm{Acceptance\, Rate} (AR) = \frac{ \textrm{no. of physically feasible solutions} }{\textrm{Total no. of solutions}} = \frac{m}{n\times n}
\end{equation}
Here, $m$ is the no. of solutions accepted after applying constrain and $n$ is no. the of samples drawn from the $t_{1AU}$ and $v_0$ distributions each.
Figure \ref{fig:flowchart} shows the flow diagram for the DBM inversion procedure that we implemented on the CME dataset and how the results of the inversion procedure are analysed.

\section{Results}
\label{RESULTS}
\subsection{Inversion Procedure Results}
\begin{figure}[h]
   \centering
   \hfill
        \subfigure{\includegraphics[width=0.32\textwidth]{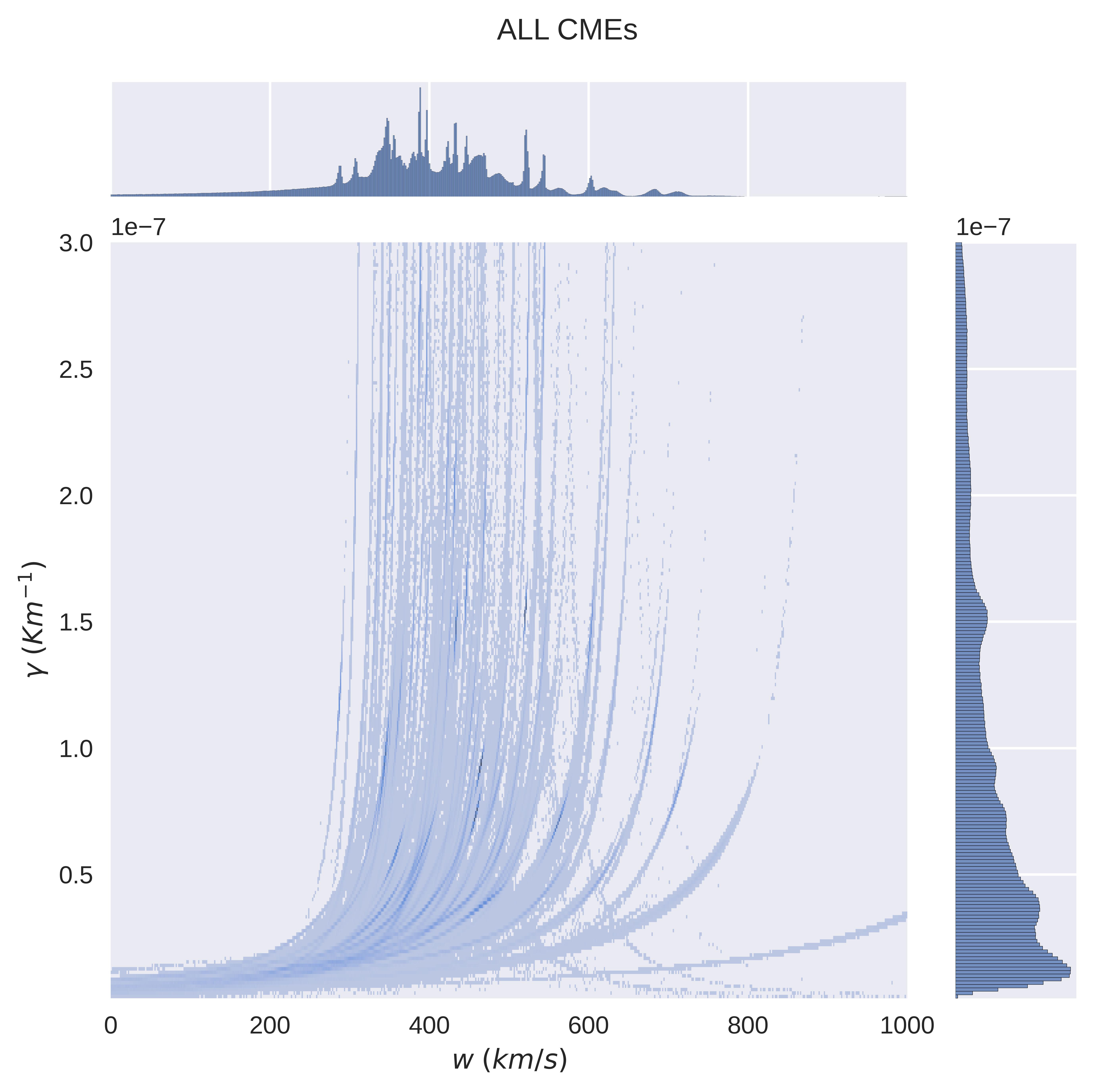}}
   \hfill
        \subfigure{\includegraphics[width=0.32\textwidth]{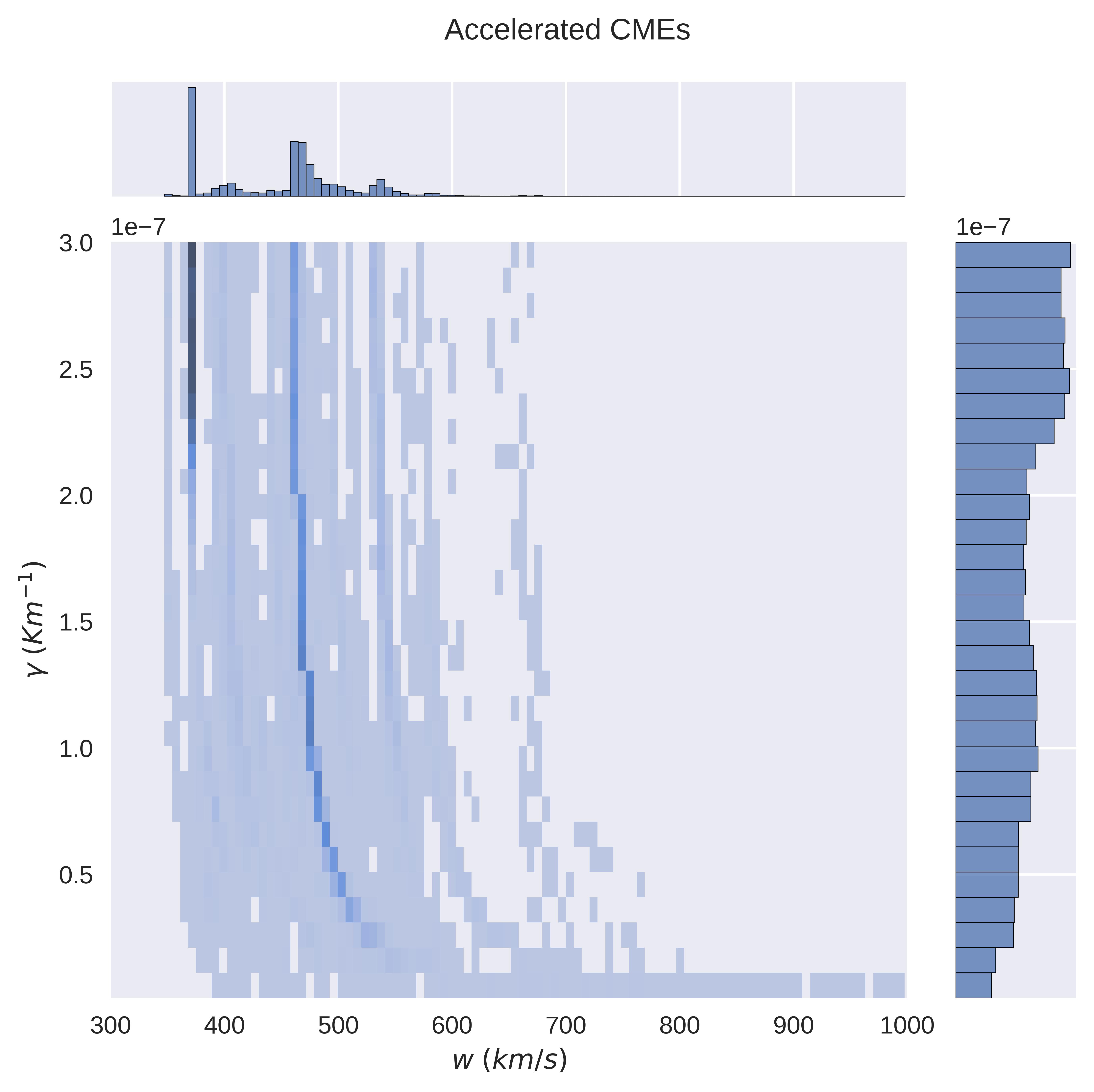}}
    \hfill
        \subfigure{\includegraphics[width=0.32\textwidth]{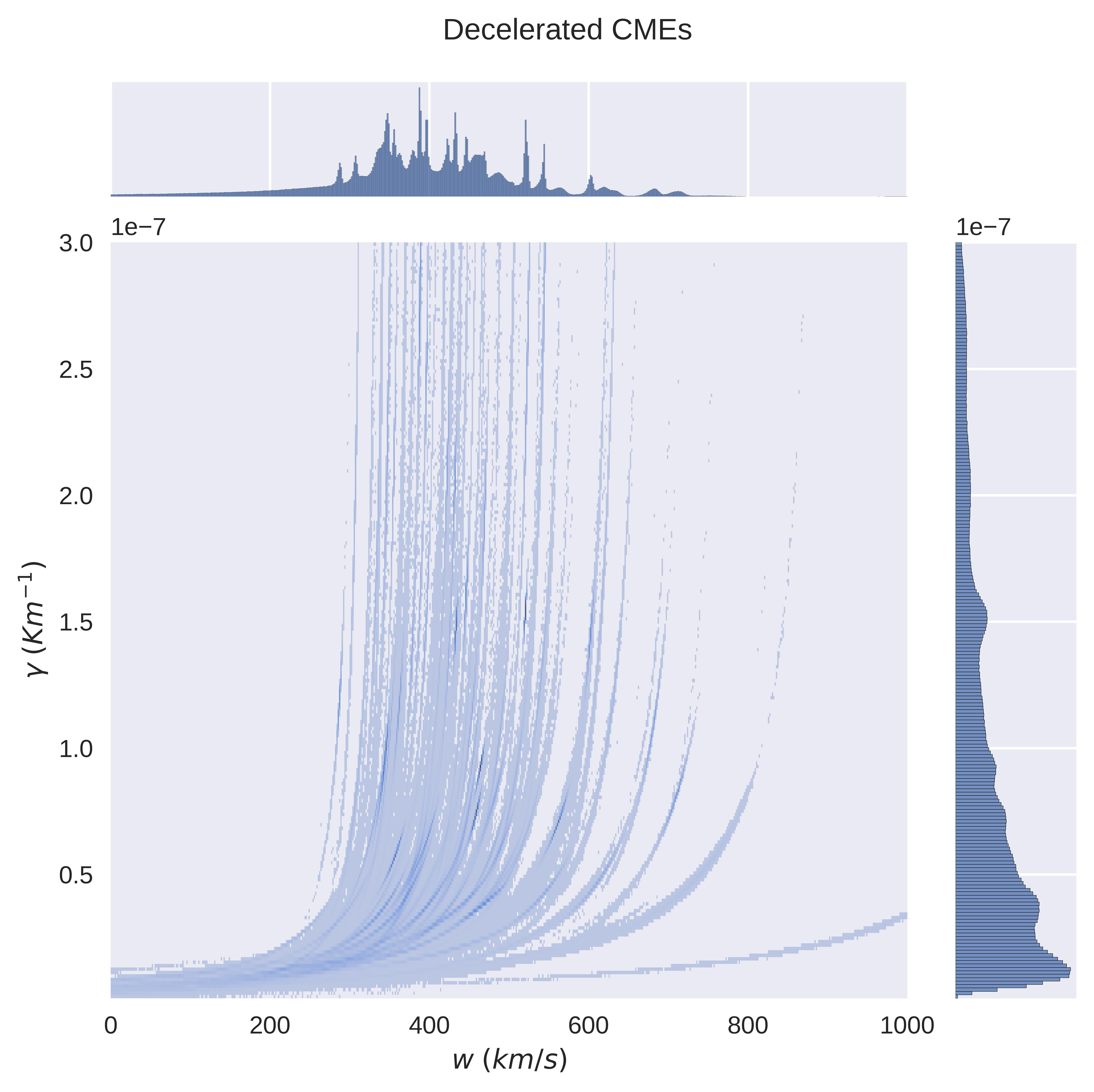}}

        \caption{Joint distribution of ($\gamma$,$w$) from the inversion procedure. Left Panel:($\gamma$,$w$) Phase space for whole dataset (3644748 values)
        Middle Panel:($\gamma$,$w$) phase space for the dataset of accelerated CMEs (25428 values)
        Right Panel:($\gamma$,$w$) phase space for the dataset of decelerated CMEs (3619320 values) }
        \label{WGALL}
   \end{figure}

The inversion procedure was performed on the entire CME-ICME pair catalogue and it turned out to be successful for 204 out of 213 events. 
At the end of the inversion procedure, we obtained $3,664,748$ possible values of $w$ and $\gamma$ that enable us to provide a statistical distribution for them. 
In fig \ref{WGALL}, the ($\gamma$,$w$) phase space for the entire ICME dataset is shown and there we can identify the trend line for a few individual CMEs.
From the DBM equation \ref{eqn:acceleration}, one can easily notice that CME either accelerate or decelerate during their propagation. 
Based on these propagation conditions, we derived two different distributions divided into accelerated and decelerated CMEs. 
Furthermore, a free DBM parameter $w$ can also be divided into two groups called the slow and fast solar wind, and therefore we can draw two more joint distributions based on solar wind speed conditions.\\

\subsection{Determining the quality of inversion process for each CME event}
It is important to note that, we claimed that the DBM inversion was successful for 204 events and therefore there should be $8,160,000$ possible values of $\gamma$ and $w$ that are more than double the numbers we have obtained from the DBM inversion procedure.
This discrepancy is due to the fact that there are many pairs for which the DBM inversion procedure is not successful or the obtained values of ($\gamma$,$w$) are discarded as they did not fulfil the constraints. 
This can also be observed in the ($\gamma$,$w$) phase space of different CME events.
Based on the density in the ($\gamma$,$w$) phase space, we label the event as "Nice Fit", "Poor Fit" and "Bad Fit". 
This labelling helps us to determine which CME event in the database follows the DBM hypothesis.
To stay consistent in the labelling procedure, we used the Acceptance Rate (AR) defined by the equation \ref{eqn:AR}.
The description for the labels is as follows.
\begin{enumerate}
    \item Nice Fit: AR $>$ 0.5; the DBM approximation is very well valid for this kind of CME event as the inversion procedure is successful for more than 50 $\%$ of the pairs. Therefore, there is a very sharp trendline in ($\gamma$,$w$) phase space. 
    \item Poor Fit: 0.25 $\leq$ AR $\leq$ 0.5; the DBM is reasonably valid as one can still see the trendline in ($\gamma$,$w$) phase space.
    \item Bad Fit: AR $<$ 0.25; the DBM approximation is less applicable for the events and it is hard to find the trend line in phase space.
\end{enumerate}

\begin{figure}[h]
    \centering
    \includegraphics{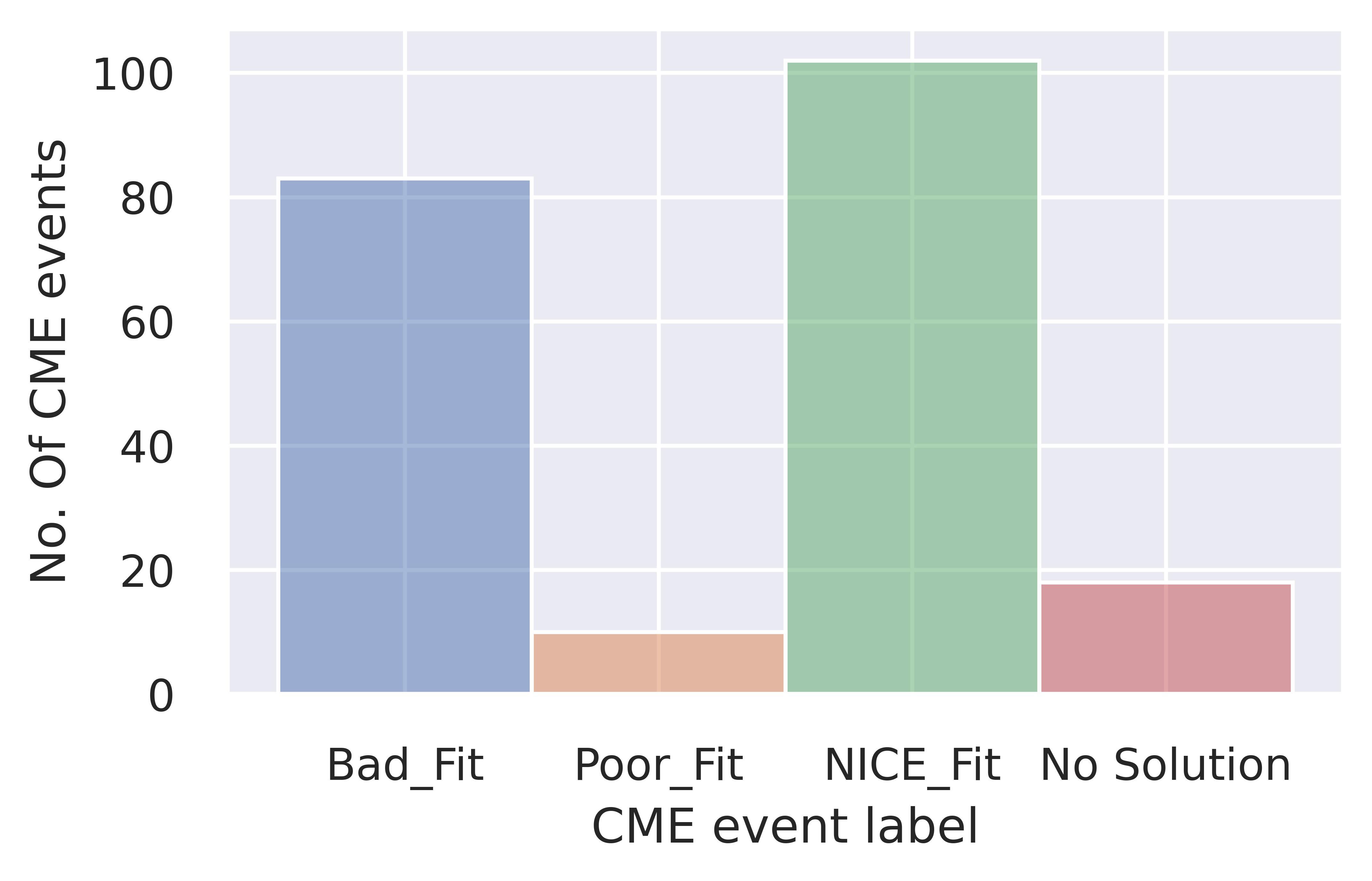}
    \caption{Histogram showing a number of events in the "Nice Fit", "Poor Fit" and "Bad Fit". }
    \label{fig:event_hist}
\end{figure}

%Initially, this labeling was carried out manually by accessing the ($\gamma$,$w$) phase space, and of course, it may be biased by an analyst.
%In the figure \ref{fig:event_hist}, number of events in each label is shown.
Figure \ref{fig:event_hist} shows event counts in each assigned label.
We want to stress here the fact that, this labeling scheme is a key point for the database that is created as a result of this work.
This labeling helps us to determine which CME event in the database follows the DBM.
Events that are flagged as "Poor Fit" or "Bad Fit" are required further investigation.
Hereafter, we only focused on the \textit{Nice Fit} events to obtain the PDFs for $w$ and $\gamma$ as it helps to improve the PDFs.
Eventually, these better statistics will lead to better accuracy in CME arrival forecasting. 
%to do: phase space for nice, poor and bad
\subsection{Relabelling the Solar Wind condition}
\begin{figure}[h]
    \centering
     \hfill
    \subfigure{\includegraphics[width=0.47\textwidth]{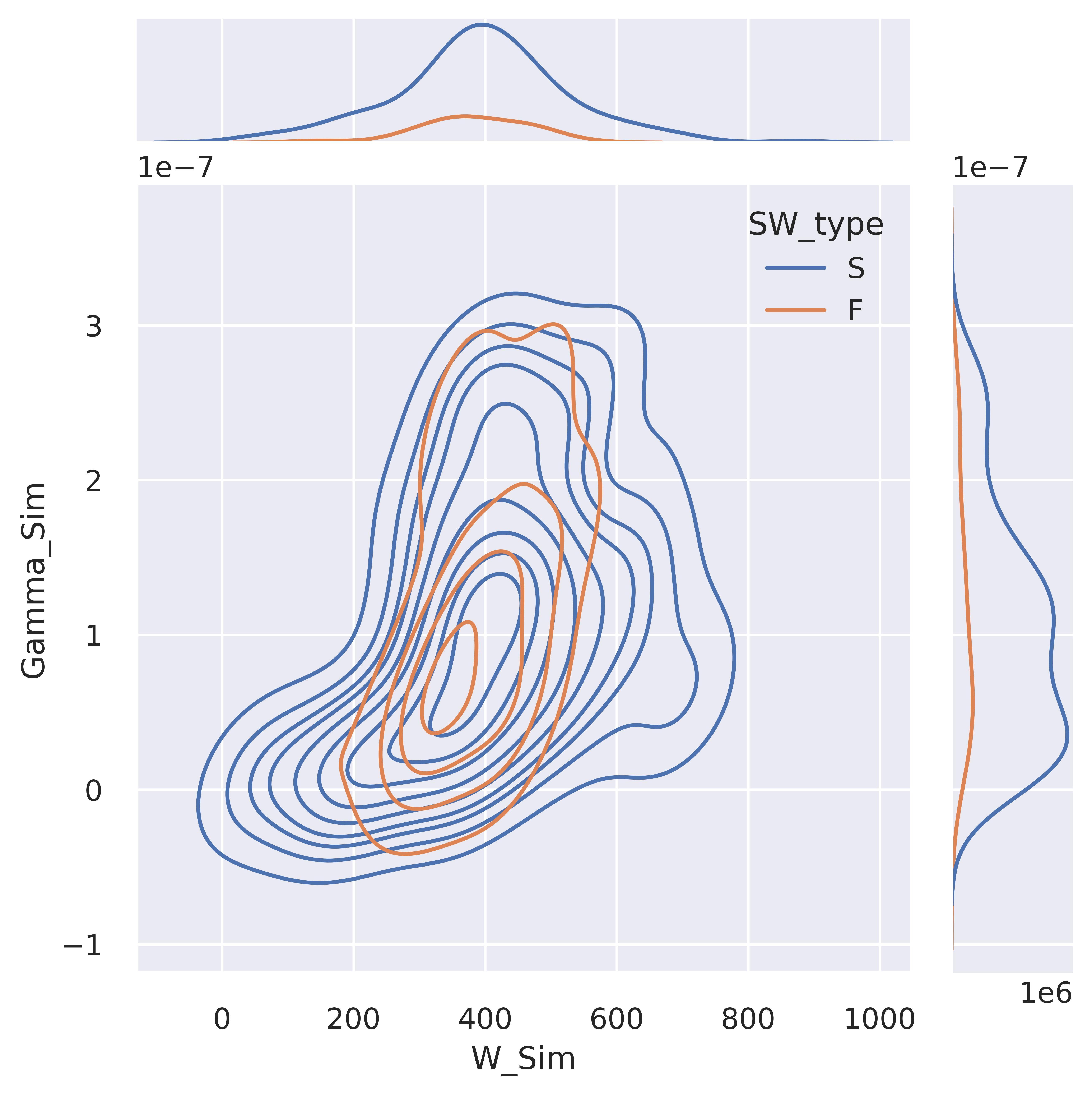}}
     \hfill
     \subfigure{\includegraphics[width=0.47\textwidth]{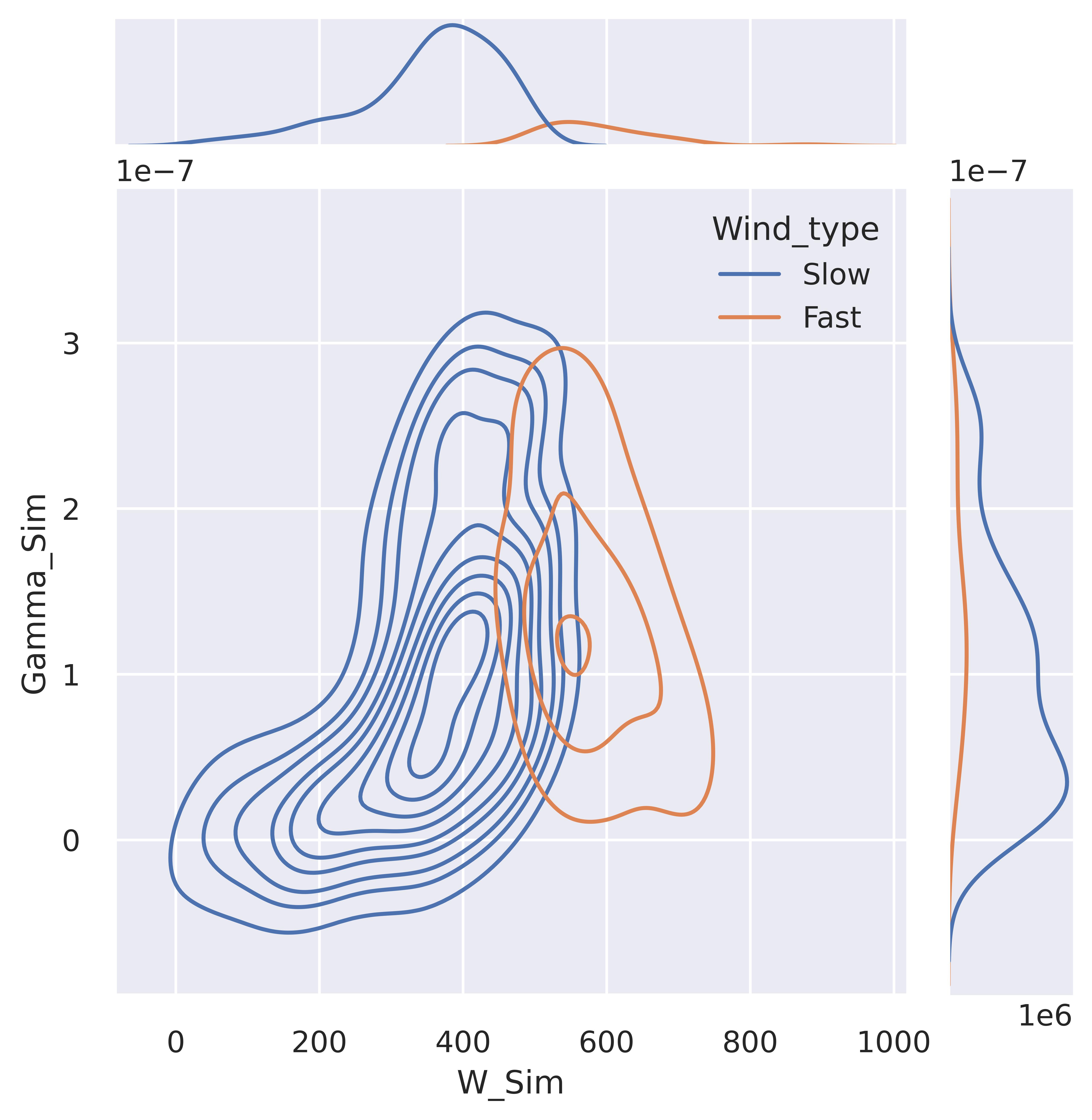}}
    \caption{($\gamma$,$w$) Phase space in different solar wind speed condition labeling scheme. On the x-axis, W\_sim shows solar wind speed obtained from a DBM inversion with a unit of km/s while on the y-axis drag (Gamma\_sim) value obtained from the inversion procedure is shown on a unit scale of km$^{-1}$  }
    \label{fig:SW_label}
\end{figure}

We found that there are only 28 CME events that are accelerating during propagation, these are around 13\% of the entire dataset, therefore statistics for accelerating CMEs are not very well resolved.
In order to find a distribution for the free DBM parameters we established a group of CME based on solar wind conditions.
A dataset that is already obtained as a part of previous work of \cite{Napoletano2022} contains information about the solar wind speed type ( See Appendix-\ref{Dataset Description} Column: SW\_type -S/F) based on the presence of Coronal Holes close to the source of a CME.
%The groups of CMEs constructed using information about the presence of coronal holes, provide two completely overlapping distributions for fast and slow solar wind speed which do not agree with our physical understanding of solar wind speed.
The group of CMEs formed based on coronal hole presence data provides two completely overlapping distributions for fast and slow solar wind speeds.
This contradicts our current physical understanding of solar wind. 
%Also, the standard deviation is high and the model cannot be useful for real-time precise and reliable space weather forecasting applications. 
Furthermore, the large standard deviation makes the model unsuitable for precise and reliable real-time space weather forecasting applications. 
Therefore we relabel the solar wind type associated with each CME by using threshold $W_{sim}\geq$ 500 km/s to discriminate the fast solar wind from the slow one. 
This threshold is similar to one that is used in \cite{Napoletano2018}. 
In figure \ref{fig:SW_label} ($\gamma$,w) phase space is shown for the two "SW\_type" and "Wind\_type" solar wind labeling. 
One important point to note here is, the tail part of any distribution in a negative region is due to the plotting style not due to the presence of any value.
Also, from now onward we focus on this new labeling scheme for solar wind speed. 

\subsection{PDF for Solar wind Speed}
\begin{figure}[h]
    \centering
     \hfill
    % \subfigure{\includegraphics[width=0.47\textwidth]{IMAGES/W_accelerated_CMEs.jpeg}}
     % \hfill
     \subfigure{\includegraphics[width=0.47\textwidth]{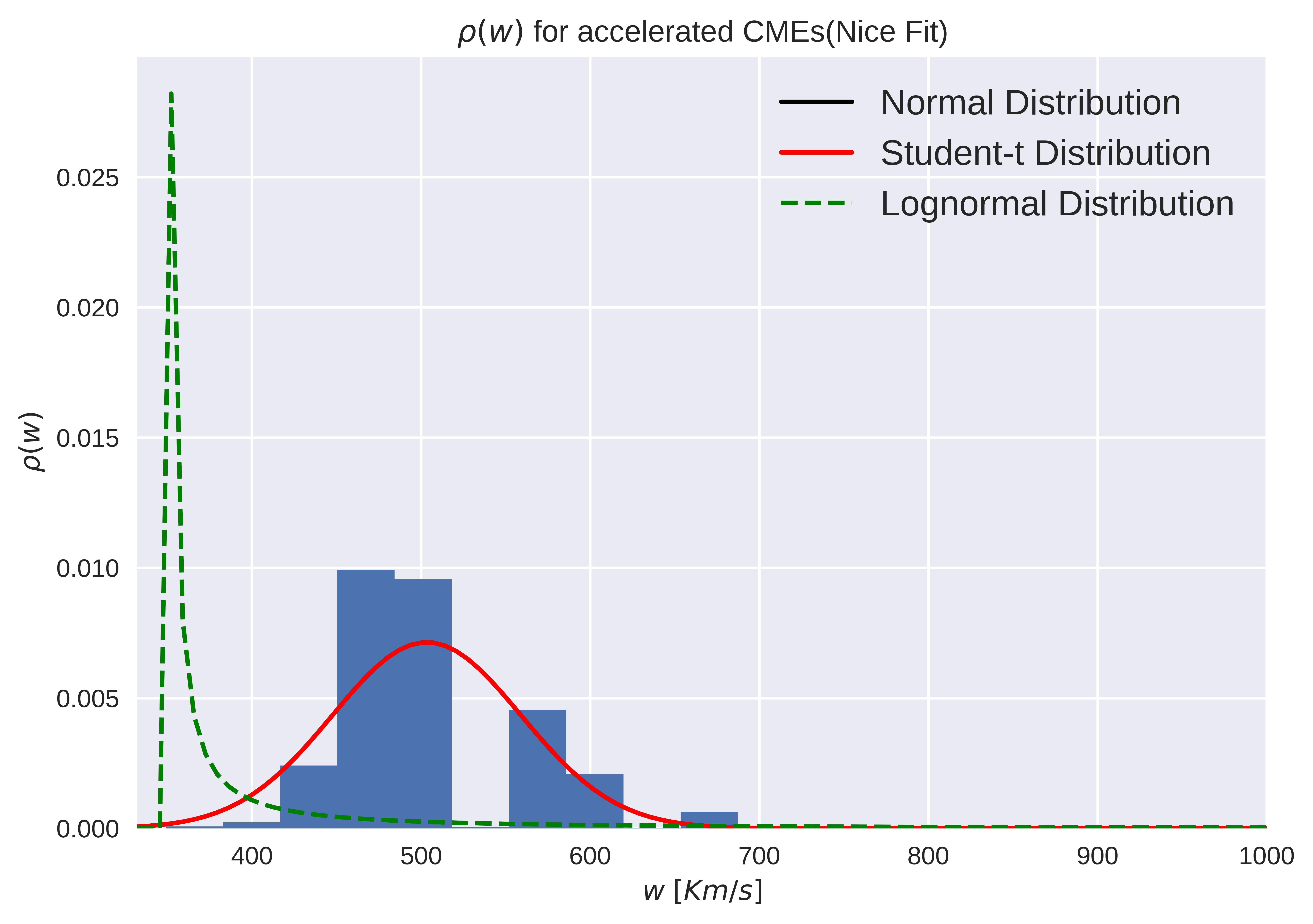}}
     \hfill
    % \subfigure{\includegraphics[width=0.47\textwidth]{IMAGES/W_decelerated_CMEs.jpeg}}
    %  \hfill
     \subfigure{\includegraphics[width=0.47\textwidth]{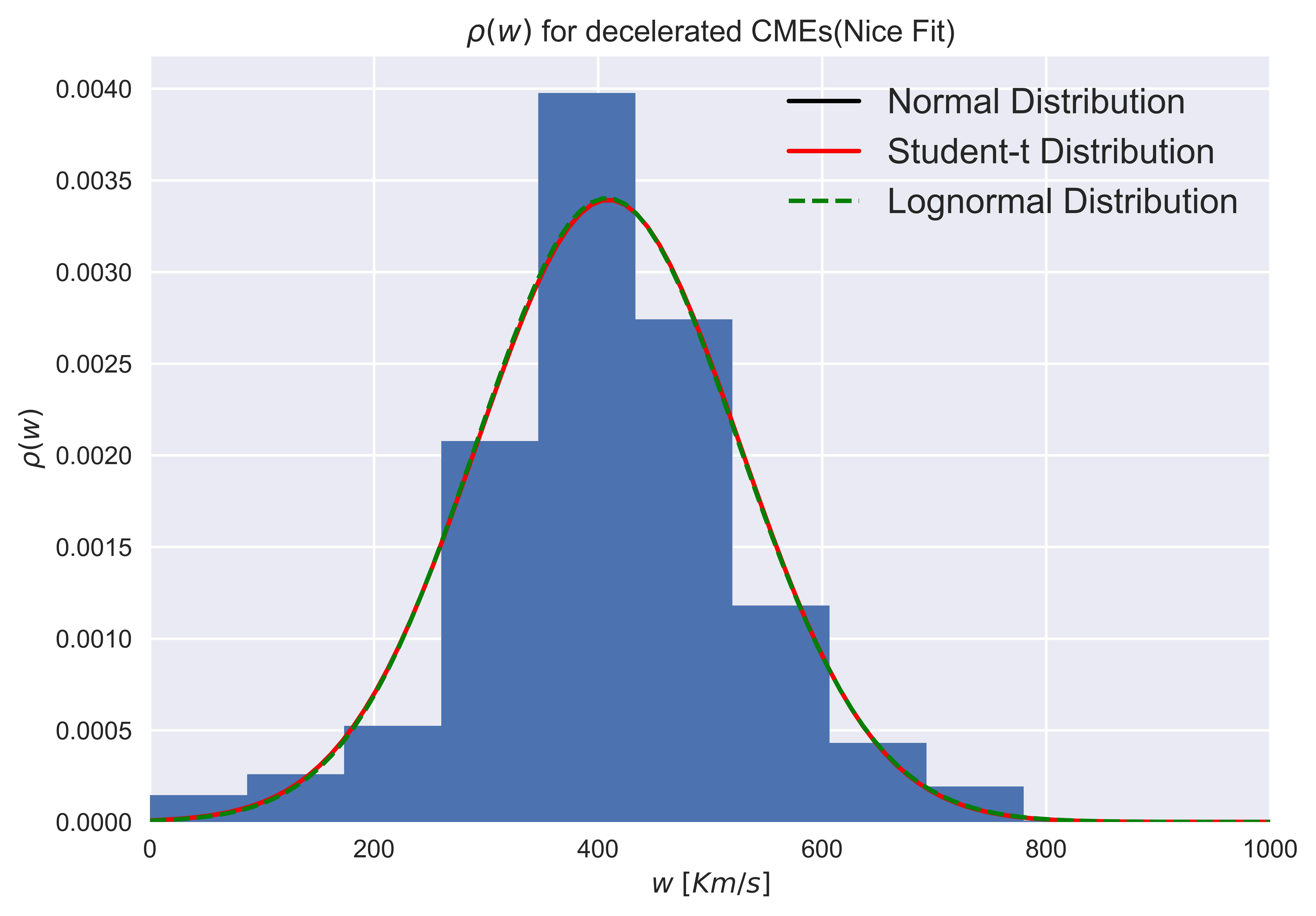}}
    \caption{Probability distribution functions for solar wind speed $w$ for accelerated and decelerated CMEs with a kernel density $\rho$ on the Y-axis. Left: $w$ PDFs for accelerated CMEs with Nice Fit label. Note that the normal and student-t distributions overlap with each other Right: $w$ PDFs for decelerated CMEs with Nice Fit label. All three distribution functions overlap with each other. The overlapping of functions is evident through RSS values. }
    \label{fig:W_acc/dec}
\end{figure}

From the joint distribution shown in figure \ref{WGALL}, we can extract a distribution function for the solar wind speed $w$. 
Here, we have fitted Gaussian, Student-t and Lognormal functions to the distribution function as these three functions returned a better fit among different PDFs available in the \textit{distfit} package \citep{taskesen_erdogan_2023_7650685}. 
In figure \ref{fig:W_acc/dec}, the histogram obtained from the dataset and fitted PDFs are shown. Here we have considered the RSS (Residual Sum of Squares) value to determine which one is the best fit. 
In most cases, all 3 distribution functions show a similar RSS value which is clear from the figure as well.
So, in the end, we concluded to select the Gaussian distribution function for the solar wind speed $w$ to be consistent with previous works of, e.g., \cite{Napoletano2018}, \cite{Dumbovic2018}, \cite{Dumbovic2021}, \cite{Napoletano2022}. 

\begin{figure}[h]
    \centering
     \hfill
    % \subfigure{\includegraphics[width=0.47\textwidth]{IMAGES/W_slow_CMEs.jpeg}}
    %  \hfill
     \subfigure{\includegraphics[width=0.47\textwidth]{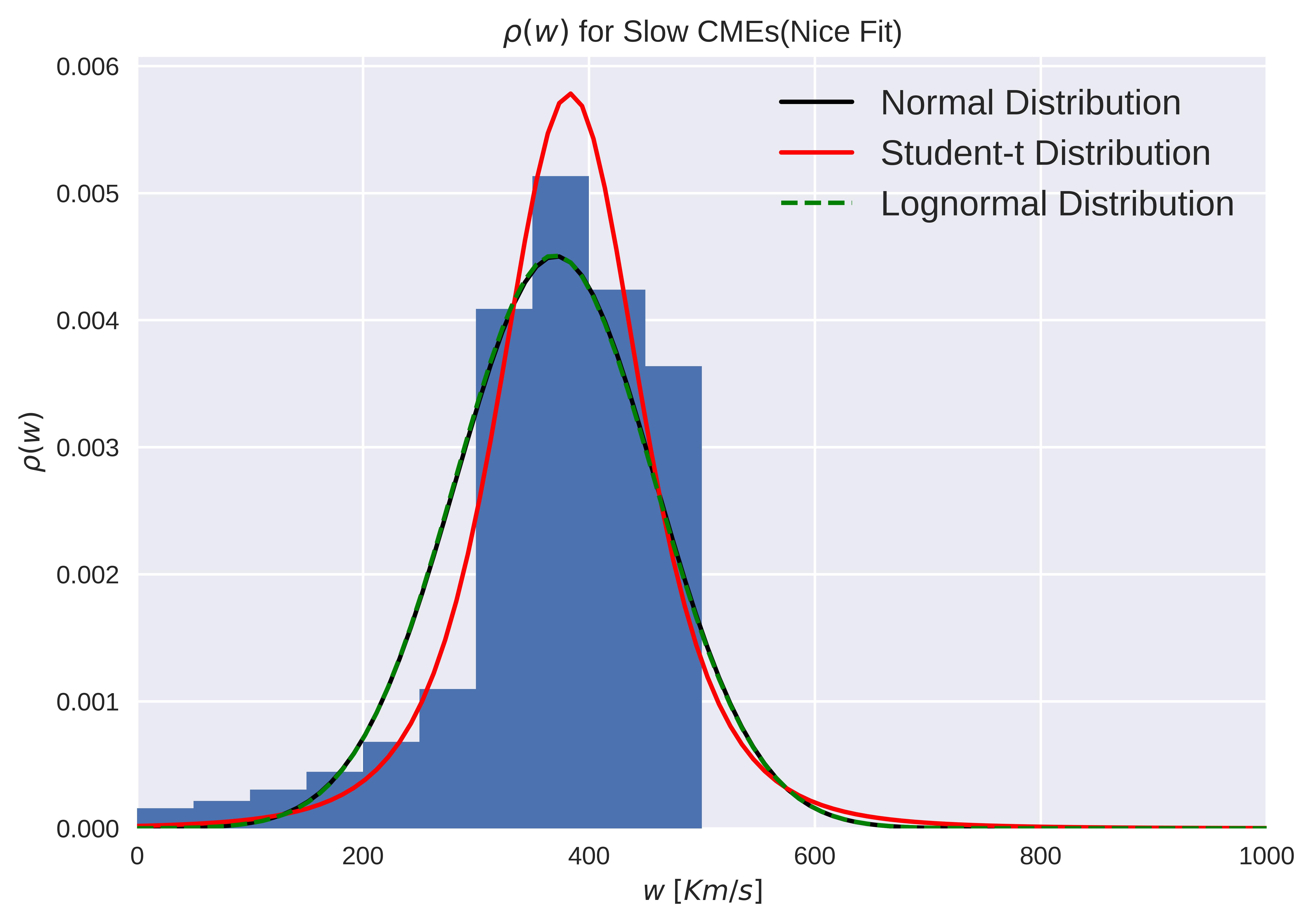}}
     \hfill
    % \subfigure{\includegraphics[width=0.47\textwidth]{IMAGES/W_fast_CMEs.jpeg}}
    %  \hfill
     \subfigure{\includegraphics[width=0.47\textwidth]{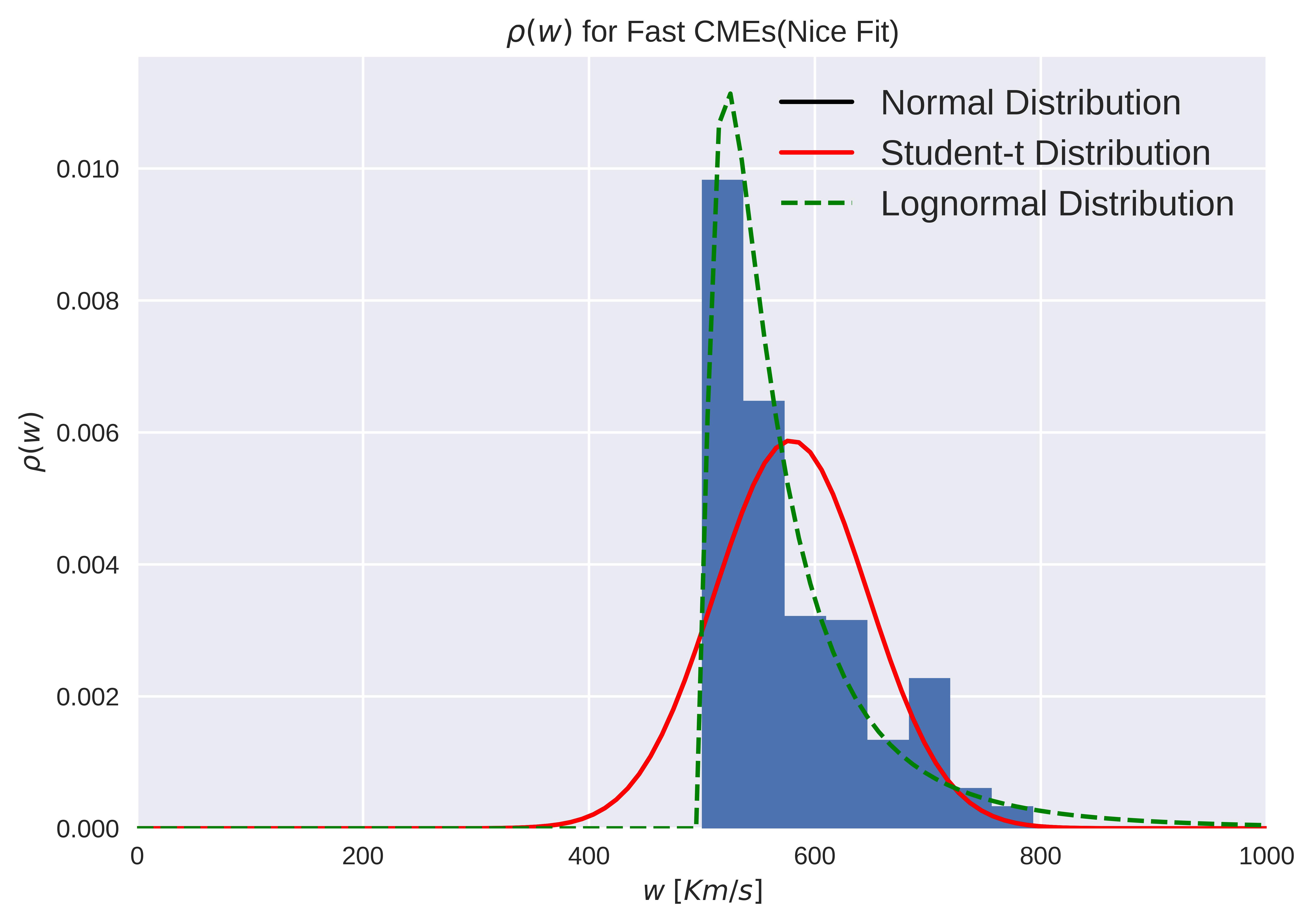}}
    \caption{Probability distribution functions for solar wind speed $w$ for slow and fast CMEs. Left: $w$ PDFs for slow CMEs with Nice Fit label. Right: $w$ PDFs for fast CMEs with Nice Fit label. (In both cases the Normal and Student-t functions overlapped with each other) }
    \label{fig:W_s/f}
\end{figure}

We categorized our dataset into Slow and Fast CMEs based on the ambient solar wind condition experienced by the CME during its propagation (using a threshold of 500km/s to separate fast and slow solar wind conditions), as described before in this section., and attempted again to fit the same three distribution functions.
Unlike the prior attempt, the fitting's RSS value is not the same for slow and fast solar wind conditions. 
For slow CMEs, the "student-t" distribution describes the best PDF while for fast CMEs the lognormal function is the most suitable PDF. 
Here, we only emphasize the fact that "student-t" and "lognormal" distributions are the best fits and are strongly biased by the hard thresholding. 
In figure \ref{fig:W_s/f} PDFs for the slow and fast solar wind conditions are shown. The parameters for the fitted distributions are reported in table \ref{table:w_param}.

\begin{table}
\caption{Parameters for the different functions used to model the Solar wind speed distribution. For the Lognormal function, tabulated values can not be used directly as average and standard deviation. The transformation from the fitting parameters to values used in the model can be done by the equation \ref{logfunc}.   }             % title of Table
\label{table:w_param}      % is used to refer this table in the text
\centering   
\begin{tabular}{lllllll}
  \hline
  CME Group & PDF & $\bar{w}[Km/s]$ & $\sigma_{w}[Km/s]$ & Args & RSS   \\
  \hline
  \hline
  Accelerated & Normal       &  503.356   &  55.848  &  -        &  0.000538  \\
              & Student's-t  &  503.356   &  55.848  &  1.526$\times10^6$ &  0.000538\\
              &  Lognormal    &  -0.407    &  3.910   &  349.009   &  0.001469\\

  \hline 
  Decelerated & Normal       &  409.168   &  117.545 &  -        & 
 0.00046  \\
              & Student's-t  &  409.168   &  117.543  &  1.479$\times10^5$ &  0.00046\\
              &  Lognormal    &  9.524    &  0.009   &  -1.328$\times10^4$   &  0.000458\\

  \hline
  Slow        & Normal       &  370.530   &  88.585  &  -        &  0.000714  \\
              &  Student's-t  &  383.169   &  64.944  &  4.101 &  0.000622\\
              &  Lognormal    &  9.784    &  0.005   &  -1.738$\times10^4$   &  0.00072\\

  \hline
  Fast        & Normal       &  579.058   &  67.871  &  -        &  0.002862  \\
              &  Student's-t  &  579.058   &  67.872  &  1.837$\times10^6$ &  0.002862\\
              & Lognormal    &  4.084    &  0.883   &  494.597   &  0.001934\\

\end{tabular}
\end{table}

\cite{Paouris2021a} has studied the same 16 CME-ICME events from the \cite{Dumbovic2018} to compare the performance of the Effective Acceleration Model (EAM) with Drag Based Ensemble Model (DBEM). 
They have also performed the inversion technique to find optimal values of solar wind speed $w$ and drag parameter $\gamma$. 
In the table \ref{tbl:wopt} below optimal values of $w$ from different studies have been shown.
It is important to note that the sample size employed in \cite{Napoletano2022} and this work is large. That helps to explain the higher value of the standard deviation.

\begin{table}[!ht]
    \centering
    \label{tbl:wopt}
    \caption{Optimal values for solar wind speed $w$ from different studies}
    \begin{tabular}{lll}
    \hline
         & \makecell{Optimal Solar wind speed \\ $w$ $[km/s]$}  & \makecell{Standard deviation\\ $\sigma_w$ $[km/s]$} \\ \hline \hline
        \cite{Dumbovic2018} & 350 & 50 \\ \hline
        \cite{Napoletano2018} (slow) & 400  & 33  \\ 
        \cite{Napoletano2018} (fast) & 600 & 66 \\
        \hline
        \cite{Paouris2021a} & 431 & 57 \\ \hline
        \cite{Calogovic2021} & 450 & 150 \\ \hline
        \cite{Napoletano2022} (slow) & 370  & 80  \\
        \cite{Napoletano2022} (fast) & 490 & 100 \\\hline
        This work (slow) & 370 & 88 \\
        This work (fast) & 579 & 68 \\ 
    \end{tabular}
\end{table}

\subsection{PDF for Drag Parameter}
\begin{figure}[h]
    \centering
     \hfill
    % \subfigure{\includegraphics[width=0.47\textwidth]{IMAGES/gamma_accelerated_CMEs.jpeg}}
    %  \hfill
     \subfigure{\includegraphics[width=0.47\textwidth]{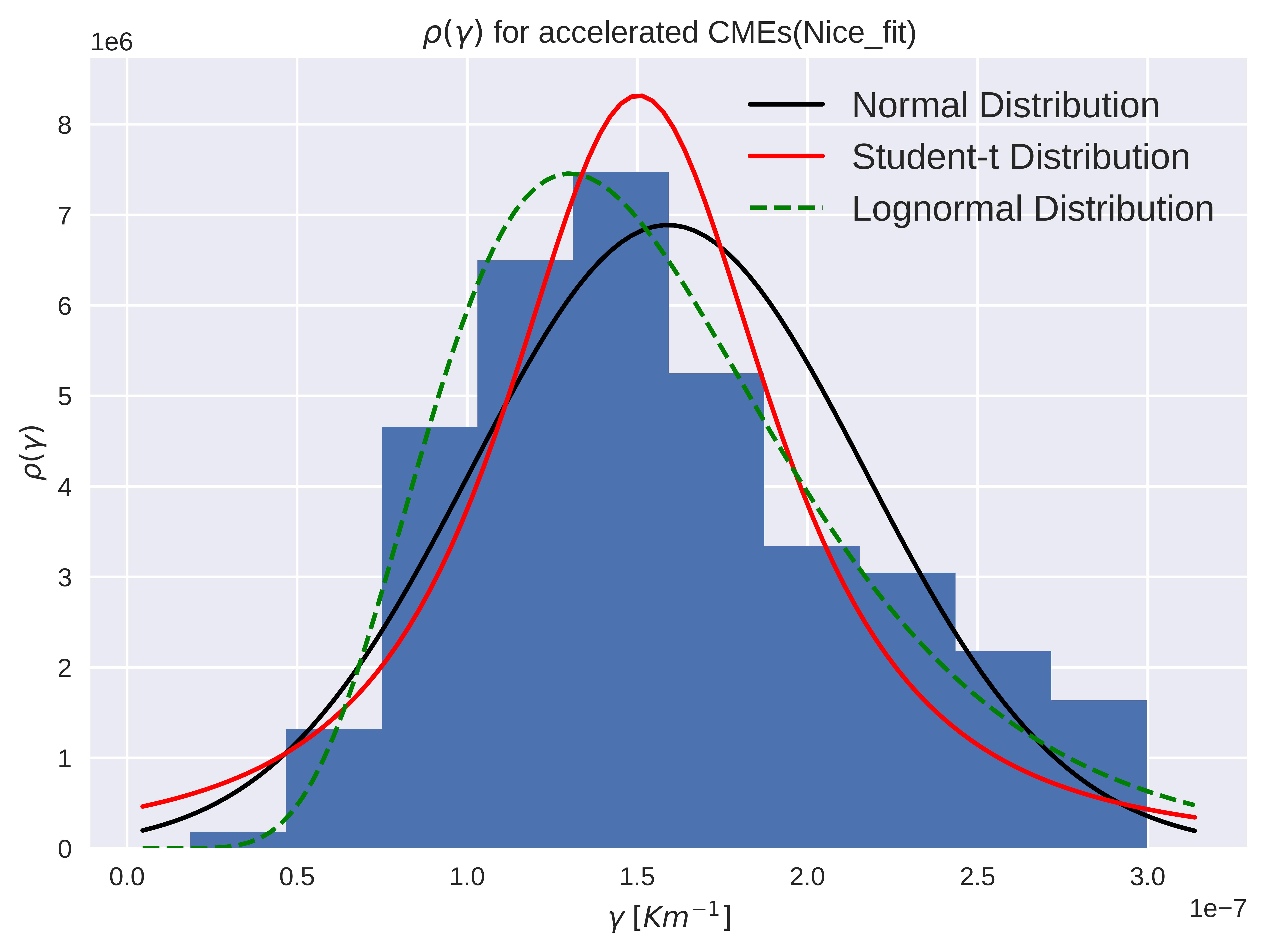}}
     \hfill
    % \subfigure{\includegraphics[width=0.47\textwidth]{IMAGES/gamma_decelerated_CMEs.jpeg}}
    %  \hfill
     \subfigure{\includegraphics[width=0.47\textwidth]{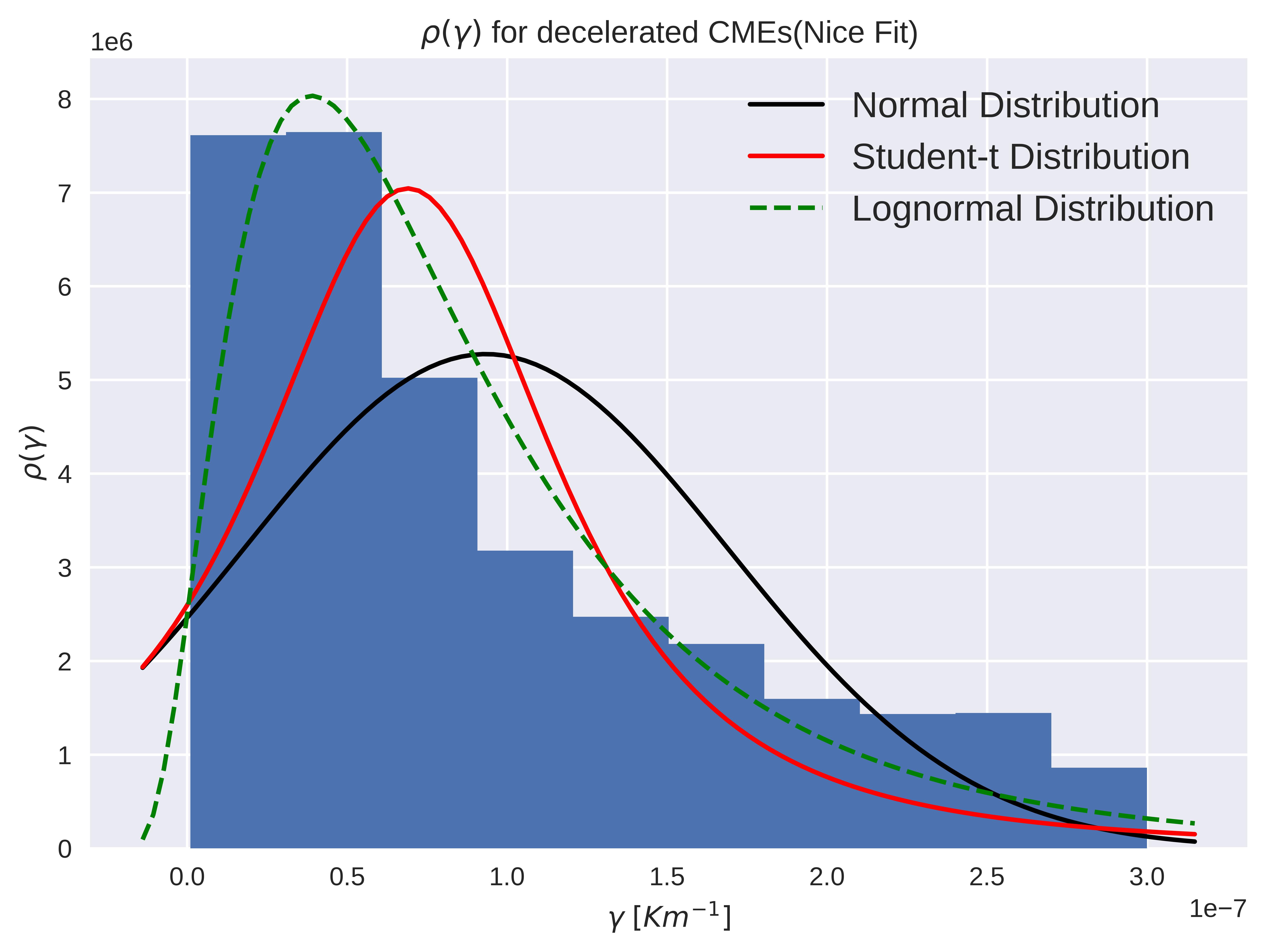}}
    \caption{Probability distribution functions for drag parameters $\gamma$ for accelerated and decelerated CMEs. Left: $w$ PDFs for accelerated CMEs with Nice Fit label. Right: $w$ PDFs for decelerated CMEs with Nice Fit label.}
    \label{fig:g_acc/dec}
\end{figure}

\begin{figure}[h]
    \centering
     \hfill
    % \subfigure{\includegraphics[width=0.47\textwidth]{IMAGES/gamma_fast_CMEs.jpeg}}
    %  \hfill
     \subfigure{\includegraphics[width=0.47\textwidth]{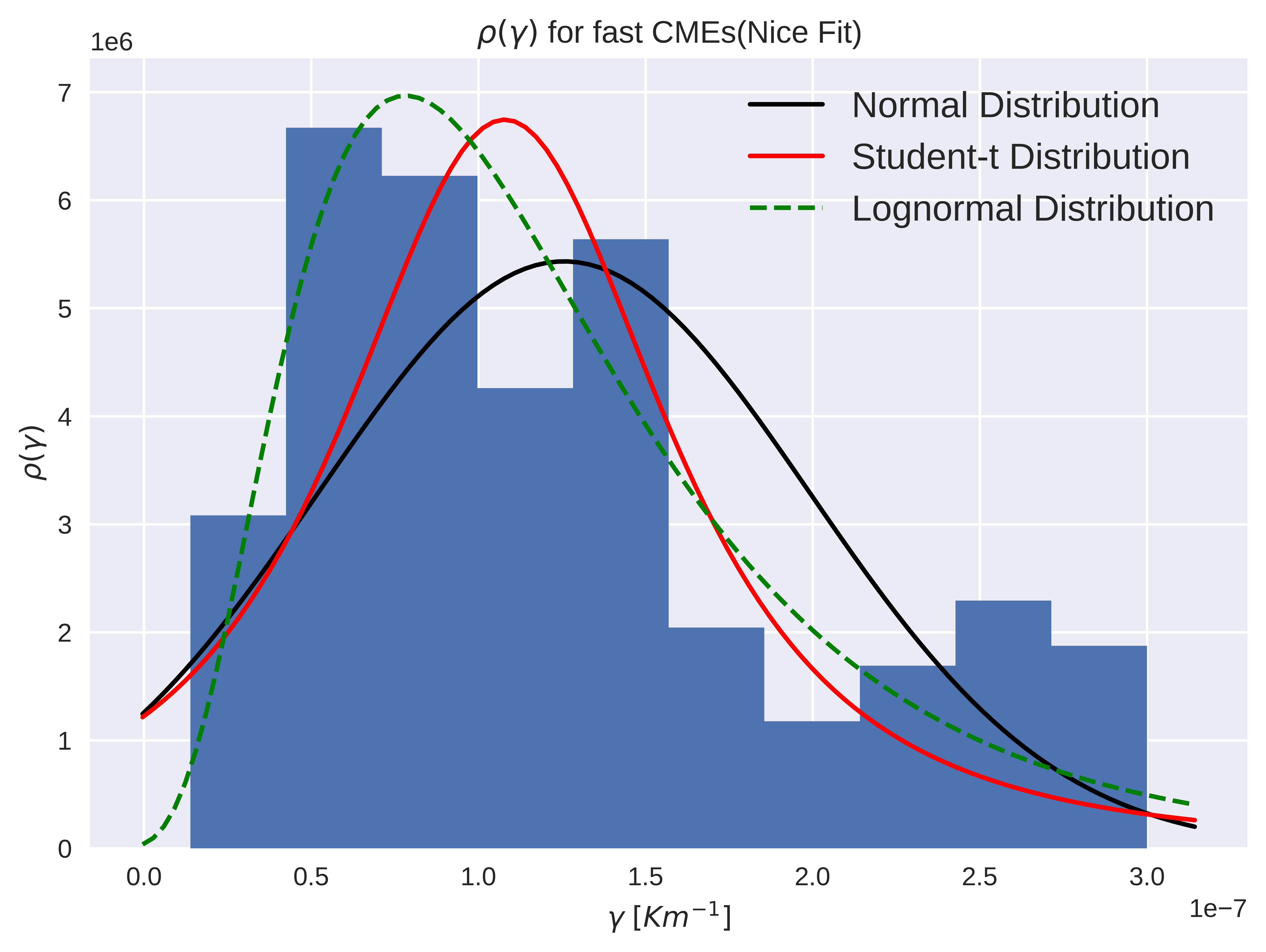}}
     \hfill
    % \subfigure{\includegraphics[width=0.47\textwidth]{IMAGES/gamma_slow_CMEs.jpeg}}
    %  \hfill
     \subfigure{\includegraphics[width=0.47\textwidth]{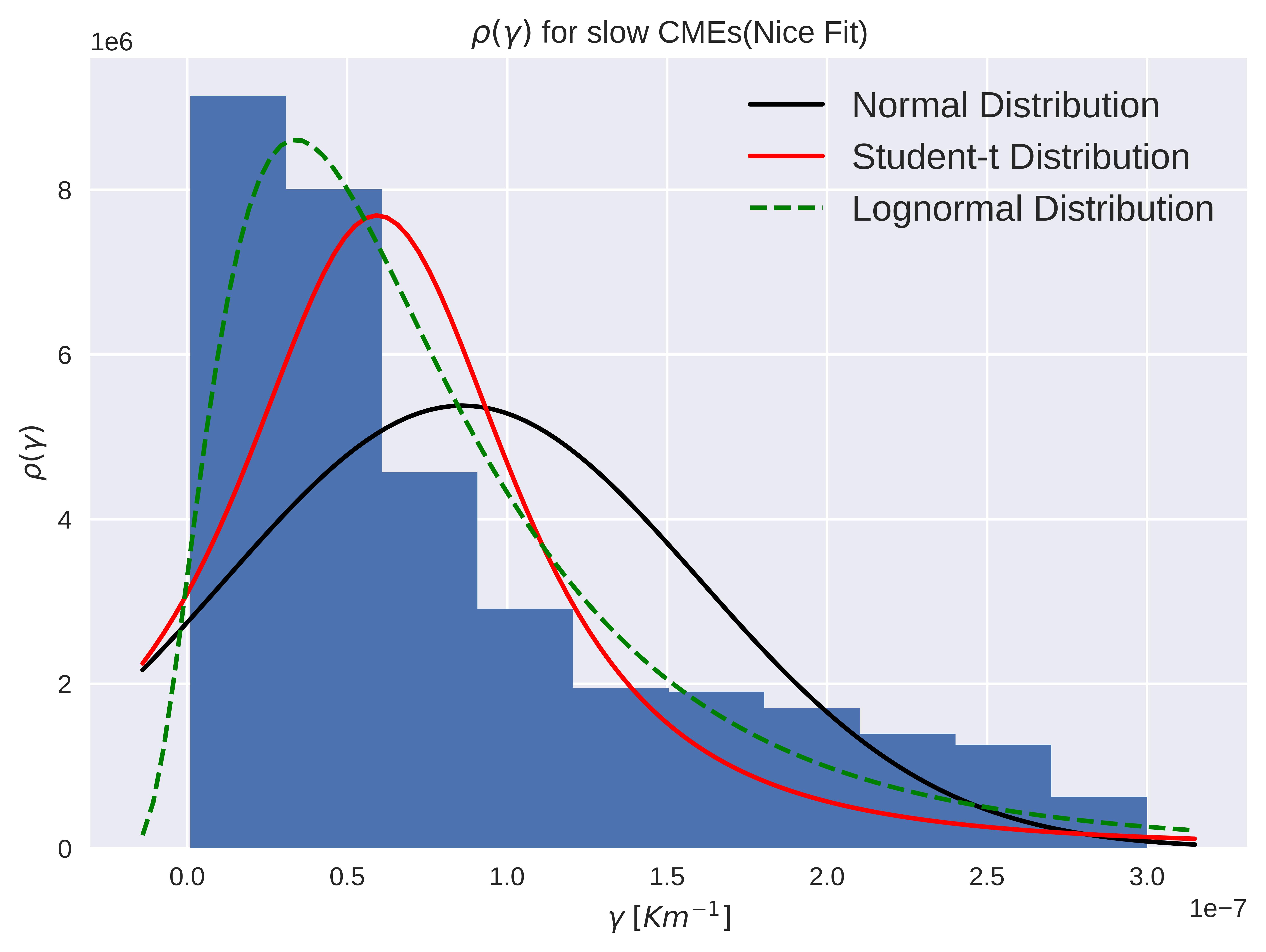}}
    \caption{Probability distribution functions for drag parameters $\gamma$ for slow and fast CMEs. Left: $w$ PDFs for fast CMEs with Nice Fit label. Right: $w$ PDFs for slow CMEs with Nice Fit label.}
    \label{fig:g_f/s}
\end{figure}

For the drag parameters, we have employed the same methods and distribution functions that we have used for the solar wind to infer the PDF. 
The RSS values obtained from the various fits are significantly different. 
%In almost every scenario, the lognormal distribution seems to be the best fit among all the distributions.
The lognormal distribution consistently emerges as the best fit among various considered distribution functions throughout a wide range of cases. 
In figure \ref{fig:g_acc/dec}, distributions fitting for the accelerated and decelerated CMEs are shown, while in figure \ref{fig:g_f/s} PDFs for slow and fast CMEs are shown. 

\begin{table}[h]
\caption{Parameter for different PDFs used to model drag parameter distribution. For the Lognormal function, tabulated values can not be used directly as average and standard deviation. The transformation from the fitting parameters to values used in the model can be done by the equation \ref{logfunc}.}             % title of Table
\label{table:g_param}      % is used to refer this table in the text
\centering   
\begin{tabular}{lllllll}
  \hline
  CME Group  & PDF & $\bar{\gamma}[Km^{-1}]$ & $\sigma_{\gamma}[Km^{-1}]$ & Args & RSS   \\
  \hline
  \hline
  Accelerated  & Normal       &  1.590$\times10^{-7}$   &  5.793$\times10^{-8}$  &  -        &  2.528$\times10^{13}$  \\
              & Student's-t  &  1.503$\times10^{-7}$   &  4.247$\times10^{-8}$  &  1.988 &  2.490$\times10^{13}$\\
              &   Lognormal    &  -15.642    &  0.354   &  -1.186$\times10^{-8}$   &  6.385$\times10^{12}$\\

  \hline 
  Decelerated & Normal       &  9.339$\times10^{-08}$ & 7.562$\times10^{-08}$ & - & 1.089$\times10^{15}$  \\
              & Student's-t  &  6.899$\times10^{-08}$ & 5.016$\times10^{-08}$ & 1.988 & 8.029$\times10^{14}$\\
              & Lognormal    &  -16.178 & 0.652 & 0.6518 & 2.723$\times10^{14}$ \\

  \hline
  Slow        & Normal       &  8.609$\times10^{-08}$ & 7.419$\times10^{-08}$ & - & 1.519$\times10^{15}$ \\
              &  Student's-t  &  5.936$\times10^{-08}$ & 4.595$\times10^{-08}$ & 1.988 & 1.010$\times10^{15}$\\
              & Lognormal    &  -16.252 & 0.658 & -2.276$\times10^{-08}$ & 4.034$\times10^{14}$\\

  \hline
  Fast         & Normal       &  1.256$\times10^{-07}$ & 7.342$\times10^{-08}$ & - & 5.319$\times10^{14}$  \\
              &  Student's-t  &  1.079$\times10^{-07}$ & 5.238$\times10^{-08}$ & 1.988 & 4.749$\times10^{14}$\\
              & Lognormal    &  -15.884 & 0.518 & -1.838$\times10^{-08}$ & 2.575$\times10^{14}$ 
\end{tabular}
\end{table}

%In table \ref{tbl:wopt}, we compare the optimal values for the solar wind speed from different studies and this work.
   
\section{Discussion and Conclusions}
\label{DISCUSSION}
% We have updated the list of CME-ICME pairs of \cite{Napoletano2022} with a DBM simulation value, PDF fitting parameters and a few other quantities for each and individual CME-ICME events.
The collection of CME-ICME pairs published in \cite{Napoletano2022} has been improved by the inclusion of DBM simulation data, PDF fitting parameters, and various other significant variables related to each individual CME-ICME occurrence.
By quantifying the success rate of the DBM inversion procedure, we were able to identify a subset of CME-ICME pairs that are well described by the DBM during their heliospheric propagation and added to the dataset the information about such categorization of CME events.
This kind of categorization delivers a lot of promise for the space weather community, as it can provide significant insights into the circumstances that make the DBM approximation fail to predict the transit time for a CME event.
On the other hand, those CME events where the DBM approximation is very valid can contribute to providing information about the model parameters $w$ and $\gamma$.
It is worth mentioning that the version of DBM we used does not consider the CME geometry, with a very simple CME front described as a spherical shell centred on the Sun.
Thus, all the CME-ICME entries that do not follow the DBM hypothesis deserve even further investigation, since we cannot tell if a 'no solution', a 'poor' or a 'bad' label actually comes from a possible error in the initial CME-ICME association, a shortage in the geometrical description of the ICME, or something happening during the ICME propagation that cannot be described by the DBM (e.g. a CME-CME interaction).
This, however, would require a thorough analysis of every single ICME and is beyond the scope of this work and may be the subject of a different work.
The revised CME-ICME collection we are presenting also includes additional details such as the solar wind speed conditions experienced by propagating CME events, more parameters about the validation of the DBM hypothesis, and information about the acceleration or deceleration mechanisms during their propagation.
The list of the improvements over the previous version published by \cite{Napoletano2022} are summarised in table \ref{dataset}. 
The revised dataset compiled and used in this work has been published at \url{https://zenodo.org/record/8063404} and a description of its columns is also provided in the appendix \ref{Dataset Description}.  

\par
As just mentioned, the subset of events where the DBM approximation holds can be employed to extract the $\gamma$ and $w$ parameters of the DBM via a Monte Carlo-like inversion procedure.
In this statistical study, we consider the uncertainties associated with the measure and the observation and incorporate them as input for the model and we only consider those CME events with more than 50\% acceptance rate in the inversion procedure. 
The reason behind this criterion is to ensure that the CME propagation is modeled by DBM with enough confidence.

\begin{figure}[h]
    \centering
    \hfill
    \subfigure{\includegraphics[width=0.47\textwidth]{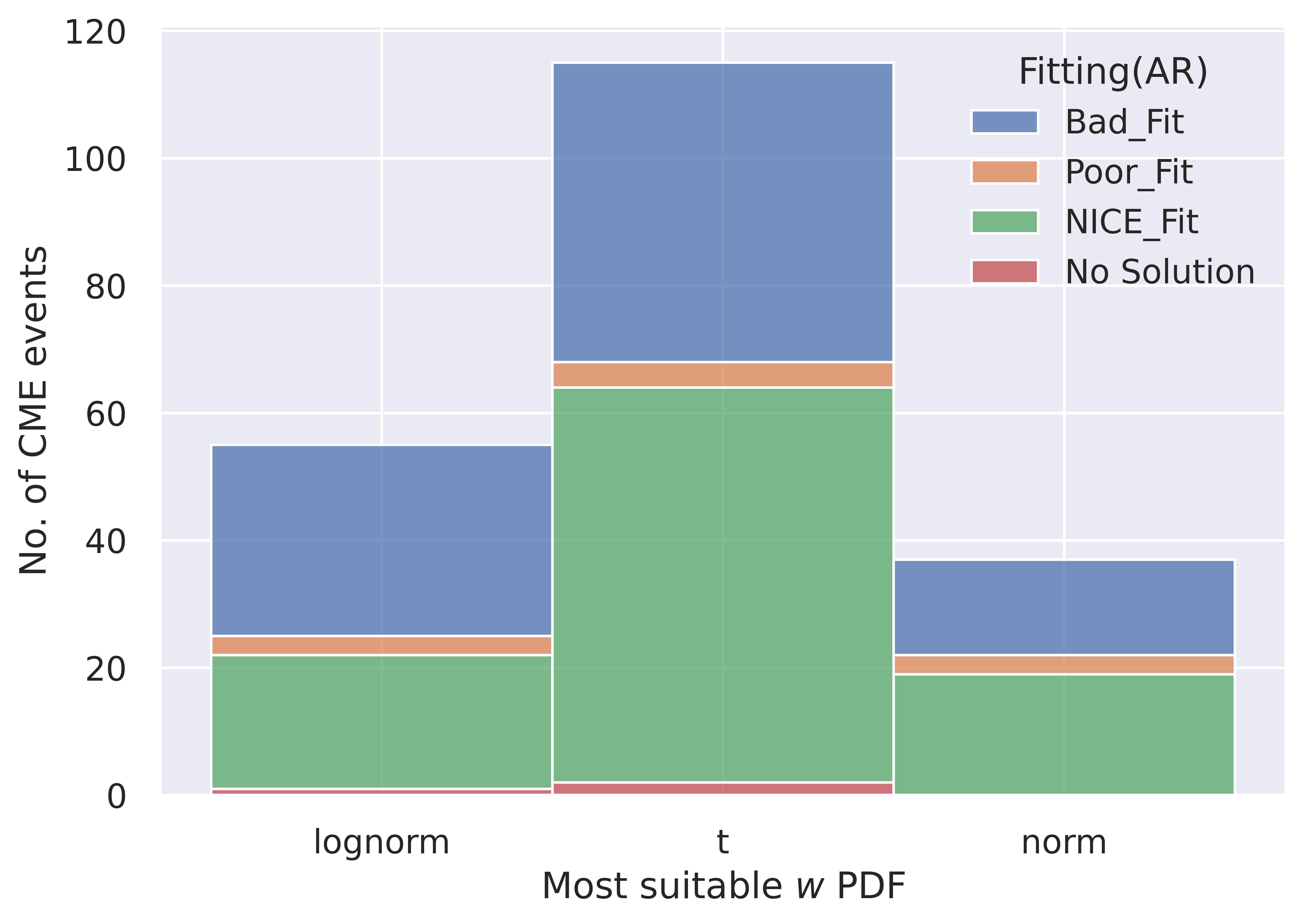}}
    \hfill
    \subfigure{\includegraphics[width=0.47\textwidth]{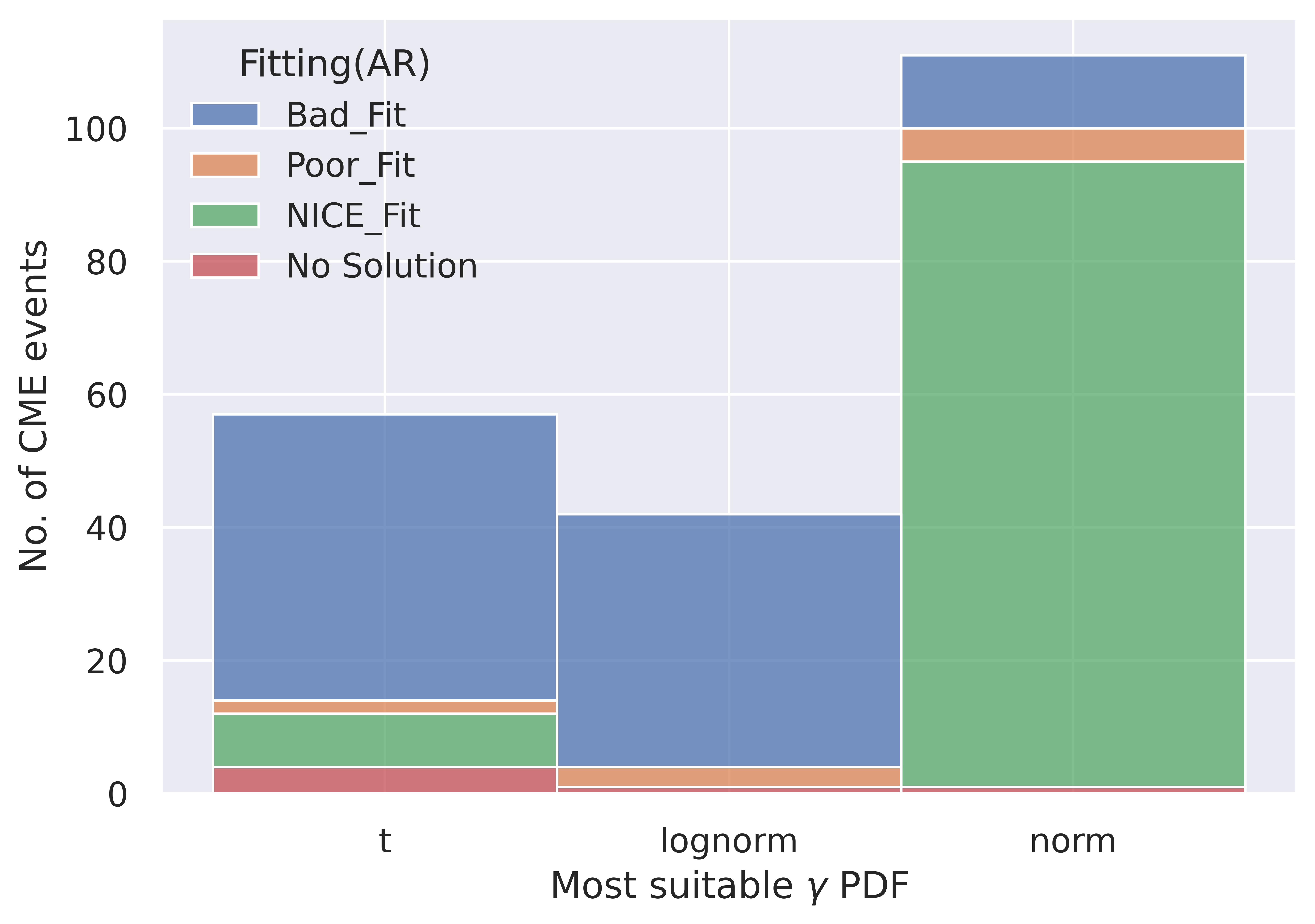}}
    \caption{Histogram illustrating most suitable PDF for solar wind speed $w$ and drag parameter $\gamma$ in single CME approach. Within each PDF, various types of CME events are stratified and effectively stacked on top of one another.   }
    \label{fig:w_hist}
\end{figure}

We have retrieved $\gamma$ and $w$ for 204 out of 213 ICMEs, which enables us to obtain robust statistics.
% Need to rephrase the following line with better wording to improve clarity.
The empirical PDF for the solar wind $w$ is modeled using two separate distributions for slow and fast solar wind conditions respectively with a threshold value of $w=500$ km/s for the fast solar wind. 
In \cite{Dumbovic2018}, \cite{Dumbovic2021}, \cite{Napoletano2018} and \cite{Napoletano2022}, a Gaussian distribution is assumed as input PDF for $w$.
Here, we have used the threshold of $w=500$ km/s for the fast solar wind speed, therefore, a normal distribution is no longer the ideal PDF. 
With this new threshold, the Student's t-distribution is the best choice for most CME events.
This latter finding is also supported by fitting PDFs for $w$ in a single CME approach. 
In Figure \ref{fig:w_hist}, a histogram of the most suitable PDFs for $w$ in individual CMEs approach is shown.
Here, the Student's t-distribution is strongly biased by the fact of hard thresholding and the RSS values of Student's-t and normal distribution are fairly comparable, we therefore prefer the Gaussian PDF for the solar wind $w$.
%Similarly, we found that the lognormal function is the most suitable PDF for the drag parameter $\gamma$. 
\par
The PDF for $\gamma$ is up for discussion from the previous works of \cite{Napoletano2018}, \cite{Napoletano2022} and \cite{Dumbovic2018},\cite{Dumbovic2021}, \cite{Calogovic2021}. 
One group employs a lognormal function, while the other group uses a Gaussian Function as input PDF.
We have tried to fit the PDF on the entire dataset and single CME events, and our study has provided light on the preference for these two different functions. 
From table \ref{table:g_param}, it is clear that lognormal distribution is the most favourable PDF as the RSS value is lower among other PDFs.  
On the contrary, when searching for the most suitable PDF in the single CME approach, the Gaussian PDF seems to be the best. 
In Figure \ref{fig:w_hist}, a histogram of the most suitable PDFs for $\gamma$ in individual CME approach is shown. 
A possible reason behind this discrepancy is the extensive dataset. 
Here, we have a used very large dataset of CME, which covers different ranges of mass and cross-sections of CME, also including almost two solar cycles’ length of CME events resulting in several kinds of solar wind density fluctuations in CME propagation.
The inclusion of all these background parameters in fitting a PDF through a dataset leads to the long-tailed lognormal function since the $\gamma$-parameter is a quantitative measure of the drag efficiency that depends on many factors such as the mass and the cross-section of the CME, and on the solar wind density \citep{Vrvsnak2013DBM}.\\

%%%%%%%%%%%%%%%%%%%%%%%%%%%%%%%%%%%%%%%%%%%%%%%
% need some closing sentences.
The refined dataset and the updated method presented in this work allowed us to explore a larger part of the $w-\gamma$ parameter space of the P-DBM model, including extreme values. 
We have investigated the possibility of $\gamma$ being a function of the ICME kinematic properties (i.e., accelerating or decelerating) or the solar wind properties (i.e., fast or slow).  
While there seems to be some difference between accelerating or decelerating ICME (see table \ref{table:g_param} and Figure \ref{fig:g_acc/dec}), the statistics need to be more robust to draw strong conclusions.
%We have established a range of potential values for slow and fast solar wind speeds that allow the space weather community to make real-time CME arrival forecasts without blindly assuming a fixed solar wind speed.
\\

% introduce possible uses of this refined database
We suggest that our result and in particular, the revised CME-ICME list will benefit the space weather community since it will provide a test bench to compare how well we can predict CME arrival time and impact.
Also, the associated information to every CME-ICME entry can help improve the accuracy and precision of other CME propagation models by including other relevant parameters.
For example, a future plan of ours is to develop a Markov Chain Monte Carlo (MCMC) approach to further constrain the PDF for $w$ and $\gamma$.
This catalogue's new entries are expected to play a relevant part in this work, promoting the convergence of Markov chains and boosting the performance of our strategy.
 
\begin{acknowledgements}
This research work has been a part of the Space Weather Awareness Training NETwork (SWATNet) project.
SWATNet has received funding from the European Union’s Horizon 2020 research and innovation programme under the Marie Sklodowska-Curie Grant Agreement No 955620.
This research has been also carried out in the framework of the CAESAR project, supported by the Italian Space Agency and the National Institute of Astrophysics through the ASI-INAF n.2020-35-HH.0 agreement for the development of the ASPIS prototype of the scientific data centre for Space Weather. 
This research has received financial support from the European Union's Horizon 2020 research and innovation program under grant agreement No. 824135 (SOLARNET).
E.C. was partially supported by NASA grants 80NSSC20K1580 ``Ensemble Learning for Accurate and Reliable Uncertainty Quantification" and 80NSSC20K1275 ``Global Evolution and Local Dynamics of the Kinetic Solar Wind".
R.E. is grateful to STFC (UK, grant No. ST/M000826/1), NKFIH OTKA (Hungary, grant No. K142987), and the Royal Society.
D.D.M. is grateful to the Italian Space Weather Community (SWICo).
\end{acknowledgements}
\paragraph{
\textit{Data Availability Statement:} The ICME catalogue built as a part of this work along with data visualisation and PDF analysis modules for implementing DBM inversion procedure can be downloaded from \url{https://zenodo.org/record/8063404} \citep{ronish_mugatwala_2023_8063404}.
}

\appendix
\section{Description of revised data set.} \label{Dataset Description}
As mentioned above, DBM inversion procedure requires initial position $r_0$, target position $r_{1AU}$, transit time $t_{1AU}$, initial speed $v_0$ and arrival speed $v_a$ to obtain $w$ and $\gamma$. 
For the purpose of this work, we have used the CME-ICME dataset from the \cite{Napoletano2022}.
This dataset contains all the required input quantities for the DBM inversion procedure.
This dataset consists of 213 CME-ICME pairs from the year 1997 to 2018, which cover a time span of two solar cycles 23 and 24.
In this dataset, information about the kinematic properties of CMEs at launch time was retrieved from the SOHO/LASCO CME Catalog \footnote{\url{https://cdaw.gsfc.nasa.gov/CME_list/}}.
While arrival time and speed of the related ICMEs have been obtained from the \cite{Richardson2010}.

\par
As mentioned in section \ref{inversion}, the uncertainty associated with different quantities is included in the inversion procedure.
SOHO/LASCO catalogue provides CME speed in the plane of sky (POS) but to make a DBM forecast more accurate de projected speed has been used in the calculation.
De projected radial speed has been obtained using equation 1 of \cite{gopalswamy2009coronal}. 
A more detailed explanation is given in appendix A2 of \cite{Napoletano2022}.
Associated solar wind speed type (column: SW\_type) for each event is hypothesised by determining the presence of a coronal hole close to the CME source region (see appendix A3 of \cite{Napoletano2022})

The Description of different columns in the database and their source work is provided in a table \ref{dataset}  

\begin{table}[!h]
    \centering
    \caption{Column description of the ICME dataset created as a part of this work}
     \begin{adjustbox}{width=\textwidth}
    \begin{tabular}{|l|l|l|l|}
    \hline
        \textit{Name } & \textit{Keyword} & \textit{Description} & \textit{Source} \\ \hline \hline
        LASCO Start & LASCO\_Start & First CME appearance in LASCO C2/C3 coronagraphs & LASCO/CDAW \\ 
        Start Date & Start\_Date & Time when CME reaches to 20 R$\odot$ & \cite{Napoletano2022} \\ 
        Arrival Date & Arrival\_Date & Estimated arrival time of ICME using insitu signatures & R \& C \\ 
        Plasma Event Duration & PE\_duration & End of ICME plasma signatures after col 3 is recorded & R \& C \\ 
        Arrival Speed & Arrival\_v & ICME arrival speed at L1 (km/s) & R \& C \\ 
        Transit Time & Transit\_time & (hrs) Computed between col 1 and col 3 & \cite{Napoletano2022} \\ 
        Transit Time Error & Transit\_time\_err & (hrs) Error associated to the start date of CME & \cite{Napoletano2022} \\ 
        LASCO date & LASCO\_Date & Most likely associated CME observed by LASCO & LASCO/CDAW \\ 
        LASCO speed & LASCO\_v & (km/s) speed correspond to the fastest moving point of CME in LASCO FOV & LASCO/CDAW \\ 
        Position Angle & LASCO\_pa & (deg)  Counterclockwise (from solar North) angle of appearance into coronographs & LASCO/CDAW \\ 
        Angular Width & LASCO\_da & (deg.) Angular expansion of CME into coronographs & LASCO/CDAW \\ 
        Halo & LASCO\_halo & If LASCO\_da is $>$270° then ’FH’ (full halo), if $>$180° ’HH’ (half halo), if $>$90° ’PH’(partial halo), otherwise ’NO & LASCO/CDAW \\ 
        De- Projected Speed & v\_r & (km/s) De-projected CME speed & \cite{Napoletano2022} \\ 
        De- Projected Speed Error & v\_r\_err & (km/s) Uncertainty of CME initial speed & \cite{Napoletano2022} \\ 
        Theta Source & Theta\_source & (arcsec) Longitude of the most likely source of CME & \cite{Napoletano2022} \\ 
        Phi Source & Phi\_source & (arcsec) Co-latitude of the most likely source of CME & \cite{Napoletano2022} \\ 
        Source POS error & source\_err & (deg.) Uncertainty of the most likely CME source & \cite{Napoletano2022} \\ 
        POS source angle & POS\_source\_angle & (deg.) Principal angle of the most likely CME source & \cite{Napoletano2022} \\ 
        Relative width & rel\_wid & (rad.) De-projected width of CME & \cite{Napoletano2022} \\ 
        Mass & Mass & (gm) Estimated CME Mass & LASCO/CDAW \\ 
        Solar Wind Type(CH) & SW\_type & Solar wind (slow, S, or fast, F) interacting with the ICME based on the presence of coronal hole near CME location & \cite{Napoletano2022} \\ 
        Bz & Bz & (nT) z-component of magnetic field at L1 and CME arrival time & R \& C \\ 
        Dst & DST & Geomagnetic Dst index recorded at CME arrival & R \& C \\ 
        Statistical de projected speed & v\_r\_stat & (km/s) Statistical de-projected CME speed, that is, $v\_r\_stat = LASCO\_v*1.027 + 41.5$ & ~ \\ 
        Acceleration & Accel. & (m/s 2) Residual acceleration at last CME observation & \cite{Napoletano2022} \\ 
        Analytical Wind & Analyitic\_w & (km/s) solar wind from DBM exact inversion & \cite{Napoletano2022} \\ 
        Analytical gamma & Analyitic\_gamma & (km$^{-1}$) drag parameter, $\gamma$, from DBM exact inversion & \cite{Napoletano2022} \\ 
        Transit Time (Simulated) & T1\_Sim & (hrs) Transit time calculated using P-DBM & This Work \\ 
        Transit Time error (Simulated) & T1\_Sim\_err & (hrs) error associated with transit time in P-DBM & This Work \\ 
        Impact Speed (Simulated) & V1\_Sim & (km/s) calculated CME arrival speed using P-DBM & This Work \\ 
        Impact Speed error (Simulated) & V1\_Sim\_err & (km/s) error associated with arrival speed in P-DBM & This Work \\ 
        Solar Wind Speed & W\_Sim & (km/s) Mean value of solar wind speed from inversion procedure & This Work \\ 
        Solar Wind Speed Error & W\_Sim\_err & (km/s) Standard deviation of solar wind speed from inversion procedure & This Work \\ 
        Gamma Simulated & Gamma\_Sim\_s & (km$^{-1}$) 's' parameter for lognormal PDF & This Work \\ 
        Gamma Error Simulated & Gamma\_Sim\_loc & (km$^{-1}$) $'loc'$ parameter for lognormal PDF & This Work \\ 
        Gamma Simulated (log) & Gamma\_Sim\_scale & $'scale'$ parameter for lognormal PDF & This Work \\ 
        Optimal Transit Time & T1\_opt & Minimally deviated transit time compared to observed one & This Work \\ 
        Optimal Impact Speed & V1\_opt & V1 correspond to T1\_opt & This Work \\ 
        Optimal W & W\_opt & W correspond to T1\_opt & This Work \\ 
        Optimal gamma & Gamma\_opt & gamma correspond to T1\_opt & This Work \\ 
        Optimal V\_r & V\_r\_opt & V\_r correspond to T1\_opt & This Work \\ 
        W CI min & W99\_min & minimum value of 99\% confidence interval for w & This Work \\ 
        W CI max & W99\_max & maximum value of 99\% confidence interval for w & This Work \\ 
        Gamma CI min & Gamma99\_min & minimum value of 99\% confidence interval for gamma & This Work \\ 
        Gamma CI max & Gamma99\_max & maximum value of 99\% confidence interval for gamma & This Work \\ 
        CME Type (V\_r\_opt) & CME\_type & CME type based on W\_sim (Accelerating/ Decelerating ) & This Work \\ 
        CME Type (V\_r) & CME\_type\_v0 & CME type based on W\_opt (accelerating: A/ decelerating D) & This Work \\ 
        Solar wind Type (Wth) & Wind\_type & Solar wind (based on threshold value) interacting with ICME & This Work \\ 
        Target distance & R1(AU) & (AU) Sun-Earth Distance at CME start date (Col2) & This Work \\ 
        Fitting & Fitting(AR) & Goodness of Inversion procedure: Nice / Poor / Bad & This Work \\ 
        Acceptance Rate & Acceptance\_Rate & Acceptance rate of inversion procedure & This Work \\ 
        Best W PDF & Best\_fit\_W & Most suitable PDF for W & This Work \\ 
        Best gamma PDF & Best\_fit\_gamma & Most suitable PDF for gamma & This Work \\ \hline
        
    \end{tabular}
    \end{adjustbox}
    \label{dataset}
\end{table}
\section{Mathematical description of Lognormal Distribution}
To find the parameters of lognormal PDF we have used a python package named \textit{distfit} \cite{taskesen_erdogan_2023_7650685} which relies on \textit{SciPy} \cite{Scipy2020}. 
The standardized form of lognormal function is given as:
\begin{equation}
    f(x,s) = \frac{1}{sx\sqrt{2\pi}} \exp(\frac{-\ln^2x}{2s^2})
\end{equation}

To shift and/or scale the above distribution function, SciPy or distfit use two more input parameters namely \textit{loc} and \textit{scale}.
With these 2 more parameters, the new function will be:
\begin{equation}
    f(x,s,loc,scale) = \frac{f(y,s)}{scale}
    \label{log1}
\end{equation}

where $y=\frac{x-loc}{scale}$. 
Suppose, a variable X is following a normal distribution with parameters $\mu$ and $\sigma$. 
Then, lognormally distributed variable Y=$exp$(X) has $\mu$ = $ln(scale)$ and $\sigma = s$.
The simplified version of formula \ref{log1} is given as follow:
\begin{equation}
    f(x,s,loc,scale) = \frac{1}{s(x-loc)\sqrt{2\pi}} exp\left[-\bigg(\frac{(ln(x-loc) - \mu)}{\sqrt{2}s}\bigg)^2\right]
\end{equation}

while a lognormal function used by \cite{Napoletano2018} is 
\begin{equation}
\label{logfunc}
    f(x,) = \frac{1}{s\sqrt{2\pi}} exp\left[-\bigg(\frac{(lnx - \mu)}{\sqrt{2}s}\bigg)^2\right]
\end{equation}

% Include data availablity statement.

%%    This version assumes use of bibtex with the jswsc.bib file being present
%%    If your bib file has a different name you need to change the following line

\newpage

\bibliography{jswsc}
   
% \end{linenumbers}
\end{document}